\documentclass[preprint,12pt]{elsarticle}

\usepackage{amssymb}
\usepackage{amsmath}
\usepackage{pifont}
\usepackage{multicol}
\usepackage{tcolorbox}
\usepackage{xcolor}
\usepackage{subcaption}
\usepackage{listings}
\definecolor{shadecolor}{gray}{0.9}
\usepackage{framed}
 {\endMakeFramed}

\usepackage{booktabs}
\usepackage{multirow}
\usepackage{color}
\usepackage{algorithmic}
\usepackage{graphicx}
\usepackage{textcomp}
\usepackage{hyperref}
\def\BibTeX{{\rm B\kern-.05em{\sc i\kern-.025em b}\kern-.08em
    T\kern-.1667em\lower.7ex\hbox{E}\kern-.125emX}}

\usepackage[linesnumbered,lined,commentsnumbered]{algorithm2e}
\usepackage{float}

\usepackage{enumitem}
\newlist{questions}{enumerate}{2}
\setlist[questions,1]{leftmargin=*, label=\textbf{RQ\arabic*.},ref=RQ\arabic*}

\usepackage{tabularx}

\newboolean{showcomments}
\setboolean{showcomments}{true} % toggle to show or hide comments
\ifthenelse{\boolean{showcomments}}
  {\newcommand{\nb}[2]{
    \fcolorbox{gray}{yellow}{\bfseries\sffamily\scriptsize#1}
    {\sf\small\textit{#2}}
   }
   
  }
  {\newcommand{\nb}[2]{}
   
  }

\newcounter{quotecount}

\usepackage[normalem]{ulem} % for \sout
\usepackage{xcolor}
\usepackage{array}

\newcommand{\ins}[1]{#1} % please insert
\newcommand{\del}[1]{} % please delete
\newcommand{\chg}[2]{#2} % please change

\newcommand{\insTwo}[1]{#1} % please insert
\newcommand{\delTwo}[1]{} % please delete
\newcolumntype{C}[1]{>{\centering\arraybackslash}p{#1}}

\newcommand{\todo}[1]{
    \ifthenelse{\boolean{showannotations}}%
    {\ifthenelse{\equal{#1}{}}{\textcolor{red}{TODO}}{\textcolor{red}{TODO:~{#1}}}}%
    {}%
}

\newcommand{\assignedto}[1]{%
    \ifthenelse{\boolean{showannotations}}%
    {\textbf{\noindent\ding{46}\textcolor{white}{~\colorbox{\assignementcolor}{Assigned to:}}~\textcolor{\assignementcolor}{#1}\\}%
    }
    {}
}

\renewcommand{\nb}[4]{
    \fcolorbox{gray}{#2}{\bfseries\sffamily\scriptsize{#1}}%
	{\sf\small$\blacktriangleright$\textcolor{#4}{\textit{#3}}$\blacktriangleleft$}%
}

\newcommand{\rem}[1]{%
    \ifthenelse{\boolean{showannotations}}%
    {\textcolor{oldtextcolor}{\st{#1}}}%
    {}%
}

\newcommand\add[1]{%
    \ifthenelse{\boolean{showannotations}}%
    {\textcolor{newtextcolor}{{#1}}}%
    {#1}%
}

\newcommand\rep[2]{%
    \ifthenelse{\boolean{showannotations}}%
    {\rem{#1}~\add{#2}}%
    {#2}%
}

\newcommand{\me}[1]{{{\small \texttt{#1}}}}

\begin{document}

\begin{frontmatter}

\setcounter{section}{0}
\setcounter{page}{1}
\setcounter{table}{0}
\setcounter{equation}{0}



%\title{Code generation of domain-specific languages with large language models}
\title{Detecting Semantic Alignments between\\Textual Specifications and Domain Models}


\author[independent]{Shwetali Shimangaud} %% Author name
\address[independent]{Independent}
%{School of Computer Science, McGill University, 3840 University, Montreal QC H3A 0E9, Canada}

\author[itis]{Lola Burgueño} %% Author name
\address[itis]{ITIS Software, Universidad de Málaga, Avda. Cervantes 2, 29071 Málaga, Spain}

\author[cordis,itis]{Jörg Kienzle} %% Author name
\address[cordis]{Cordis, Avda. Juan Lopez de Peñalver 17, P3A 2 Izq., 29590 Málaga, Spain}

\author[navcanada]{Rijul Saini} %% Author name
\address[navcanada]{NAV CANADA, 151 Slater St. Suite 120, Ottawa, Ontario, Canada K1P 5M6}

%% Abstract
\begin{abstract}

% 1.) Introduction. In one sentence, what’s the topic?
  %Deriving domain models from textual descriptions is a challenging and error-prone task, especially for novice modelers.
\textbf{Context:} Having domain models derived from textual specifications has proven to be very useful in the early phases of software engineering. However, creating correct domain models and establishing clear links with the textual specification is a challenging task, especially for novice modelers.\\
  % 2.) State the problem you tackle. What’s the key research question?
  %Sometimes, these models do not faithfully capture the intended semantics of the textual description, leading to inconsistencies, ambiguities, or missing elements. % in the final domain model.
  % 3.) Summarize (in one sentence) why nobody else has adequately answered the research question yet.
  %Until now, there has been no automated approach that can systematically identifies misalignments between textual descriptions and their corresponding domain models.
  %Until now, there has been no automated approach that can 
  %generate a perfect domain model from a textual specification, nor 
  %systematically identify model elements that align with the textual description %
  %(i.e., are correct) or contradict it (i.e., are incorrect).
  % 4.) Explain, in one sentence, how you tackled the research question.
\textbf{Objectives:} We propose an approach for determining the alignment between a partial domain model and a textual specification.\\
  % 5.) In one sentence, how did you go about doing the research that follows from your big idea. Did you run experiments? Build a piece of  software? Carry out case studies? This is likely to be the longest sentence
  %Our method generates an artificial natural language specification by traversing the partial domain model. We then employ a matching mechanism to associate these generated sentences with potential counterparts in the textual specification that address the same semantic concepts. To assess alignment, we leverage a large language model (LLM) to determine whether the matched sentences convey equivalent meanings. Each model element in the partial domain model is classified as aligned, misaligned (i.e., contradicts), or unclassified.
\textbf{Methods:} To this aim, we use Natural Language Processing techniques to pre-process the text, generate an artificial natural language specification for each model element, and then use an LLM to compare the generated description with matched sentences from the original specification.
  %Our method generates an artificial natural language specification from the partial domain model and uses Natural Language Processing techniques to find potential matches between these sentences and sentences coming from the specification. Finally, the approach prompts a large language model (LLM) to determine whether the generated sentences are semantically equivalent to the textual specification.
Ultimately, our algorithm classifies each model element as either \textit{aligned} (i.e., correct), \textit{misaligned} (i.e., incorrect), or \textit{unclassified} (i.e., insufficient evidence). Furthermore, it outputs the related sentences from the textual specification that provide the evidence for the determined class.\\
  %
  %We have evaluated our approach on 30 examples from the literature from diverse domains, each consisting of a textual specification and a reference domain model expressed using class diagrams. From each reference model we also derived an erroneous model by applying systematic semantic mutations, ending up with a final dataset of 60 examples. 
\textbf{Results:} We have evaluated our approach on a set of examples from the literature containing diverse domains, each consisting of a textual specification and a reference domain model, as well as on models containing modeling errors that were systematically derived from the correct models through mutation.
  % 6.) As a single sentence, what’s the key impact of your research?
Our results show that we are able to identify alignments and misalignments with a precision close to 1 and a recall of approximately 78\%, with execution times ranging from 18 seconds to 1 minute per model element.\\
\textbf{Conclusion:} Since our algorithm almost never classifies model elements incorrectly, and is able to classify over 3/4 of the model elements, it could be integrated into a modeling tool to provide positive feedback or generate warnings, or employed for offline validation and quality assessment.
%For detecting the misalignments, the precision is also \textcolor{red}{xx} and the recall \textcolor{red}{xx}.
\end{abstract}


% 3 to 5 bullet points, each a maximum of 85 characters including spaces
%\item Discusses the challenges of generating code for constraint and query domain-specific languages (DSLs) using large language models (LLMs).
%\item Novel method to identify (mis)alignments between textual descriptions and domain models 
%\item Evaluation performed on publicly available benchmark of 30 textual requirements and domain models
%\item Result show close-to-perfect precision and a recall of over 75\% 

%% Keywords
\begin{keyword}
% Keywords: 1 to 7 keywords
Large language models \sep domain models \sep textual requirements \sep (mis)alignment detection



\end{keyword}

\end{frontmatter}



%% Use \subsubsection, \paragraph, \subparagraph commands to 
%% start 3rd, 4th and 5th level sections.
%% Refer following link for more details.
%% https://en.wikibooks.org/wiki/LaTeX/Document_Structure#Sectioning_commands

\section{Introduction}
\label{sec:intro}

Textual specifications provide detailed descriptions of a system's desired behavior and constraints in natural language. On the other hand, domain models\footnote{Sometimes also called concept models or conceptual schemas.} are abstract, often visual representations of the key concepts and relationships relevant to the system under construction. Domain models have been shown to be very useful for purposes, e.g. communication and checking the completeness of requirements~\cite{Arora19}. However, for novice modelers, building domain models and/or understanding how to establish clear, structured links between the two kinds of artifacts is challenging.

Despite numerous efforts in recent years to (semi-)automatically generate domain models from problem descriptions~\cite{yang2022towards, saini2022machine, robeer2016automated, francuu2011automated,10.1145/3652620.3687807, prokop2024enhancing, chaaben2023towards,Silva25} as well as the associated traces (also known as links or relations) between the text and the model elements, the resulting models still require human validation before they can be reliably used in a model-driven development process.

As a result, it is essential for software developers to acquire strong modeling skills. The latest curricula from the Association for Computing Machinery (ACM) recognize modeling as a fundamental competency in software engineering education, emphasizing the importance of students becoming proficient in creating, analyzing, and refining software models~\cite{acmSEcurriculum, acmCScurricula}.
Unfortunately, theoretical knowledge of modeling languages and semantics alone is not sufficient. Becoming a skilled modeler demands extensive practice---a skill that is challenging to develop.

One way to help modelers improve their modeling skills is by indicating to them the model elements that they modeled correctly as well as identifying the mistakes they made together with a rationale. This, however, is a challenging task, since modeling is a creative activity and there is not a single correct domain model for a given situation. 

In this paper, we present an approach to identify semantic alignments and misalignments between textual specifications (i.e., requirements) and domain models which can be either complete or only partially complete (a.k.a., partial domain models) by means of LLMs. A piece of specification and an excerpt of a domain model are semantically aligned if they convey the same information. In contrast, they are misaligned if they convey contradictory information. %Note that if a piece of specification and an excerpt of domain model convey different information, they are neither aligned nor misaligned.

This paper is structured as follows. Section~\ref{sec:motivation} presents the motivation of our work and a running example. Section~\ref{sec:approach} outlines the main components of our approach and explains details of the LLM-based classification algorithm. Section~\ref{sec:evaluation} presents our evaluation and showcases its results. Section~\ref{sec:discussion} discusses our approach in the broader context of model-driven development. Section~\ref{sec:rw} presents the related work. Finally, Section~\ref{sec:conclusions} outlines the conclusions and future work.

\section{Motivation and Running Example}
\label{sec:motivation}

\begin{figure}
    \centering
    \includegraphics[width=0.8\columnwidth]{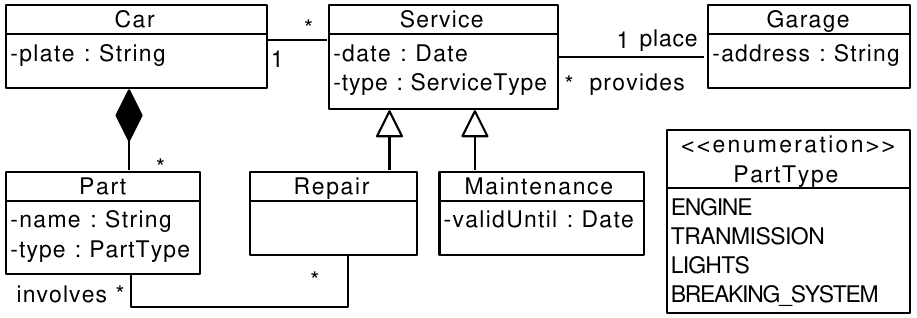}\vspace{-2mm}
    \caption{Running Example: Car Service Domain Model}
    \label{fig:carMaintenanceModel}
    \vspace{-4mm}
\end{figure}

Figure~\ref{fig:running_example} shows a short textual requirement specification that describes a car maintenance system, which we are using to illustrate our approach throughout the paper\footnote{At this point, please, ignore the fact that part of the text is in bold and underlined.}.
Based on this textual specification, a modeler might create a domain model such as the one shown in Figure~\ref{fig:carMaintenanceModel}.

%\vspace{2mm}
\begin{figure}[h]
%\noindent\fbox{%
    \centering
    \small
    \fbox{
    \parbox{0.95\columnwidth}{
    \textbf{\textcolor{gray}{Specification: }} 
    \textcolor{red}{$^1$}We have a \textbf{garage} that \underline{offers} two \textbf{types} of \textbf{services} for \textbf{car}s: \textbf{repairs} and \textbf{maintenance}.
    \textcolor{red}{$^2$}For each \textbf{car} that comes to the \textbf{garage}, the \textbf{first} \textbf{thing} to do \underline{is} to register \textbf{its} \textbf{plate} \textbf{number}.
    \textcolor{red}{$^3$}For each \textbf{service} \underline{provided}, we record the \textbf{date} and the \textbf{type} of \textbf{service}.
    \textcolor{red}{$^4$}When it \underline{comes} to \textbf{repairs}, we also note which \textbf{car} \textbf{part} was \underline{fixed} -- whether it is the \textbf{engine}, \textbf{transmission}, \textbf{lights}, or braking \textbf{system}.
    \textcolor{red}{$^5$}For \textbf{maintenance} \textbf{services}, we need to store the \textbf{date} until which the \textbf{service} is valid.
    \textcolor{red}{$^6$}Note that each \textbf{service} \underline{happens} in a \textbf{specific} \textbf{garage}, and every \textbf{garage} \underline{has} its \textbf{own} \textbf{address}.
    }
    }
    \caption{Running Example: Car Service Specification}
    \label{fig:running_example}
\end{figure}
%}
%\vspace{2mm}

In this context, our approach could be used within a modeling tool for different purposes. For example, it could be used to guide the modeler during the modeling activity. To this aim, the modeling tool could highlight the model elements that are misaligned with the textual specification as incorrect. If the user selects the misaligned model element, the tool would display the corresponding misaligned sentence(s) from the textual specification. Similarly, when our approach determines a model element to be aligned with the textual specification, the modeling tool would highlight the element as correct. That way, especially novice modelers would feel more confident about these model elements and as a result can focus on the rest of the model.

While the obvious use of our approach is to assist the modeler during the creation of the domain model, it can also be used after the creation of the model to automatically establish traceability links between the textual specification and the model elements. Especially in a model-driven engineering context, where domain models constantly evolve and are typically refined into architectural and design models and then into code, our approach can significantly ease the burden of establishing and maintaining traceability of textual requirements to models and code.

\section{Approach}
\label{sec:approach}

This section presents a detailed description of the proposed approach. First, we outline the overall architecture, offering a high-level view of the main components and how they interact. Then, we describe each module's functionality.

\subsection{Overview}

\begin{figure}[ht]
    \centering
    \includegraphics[width=\columnwidth]{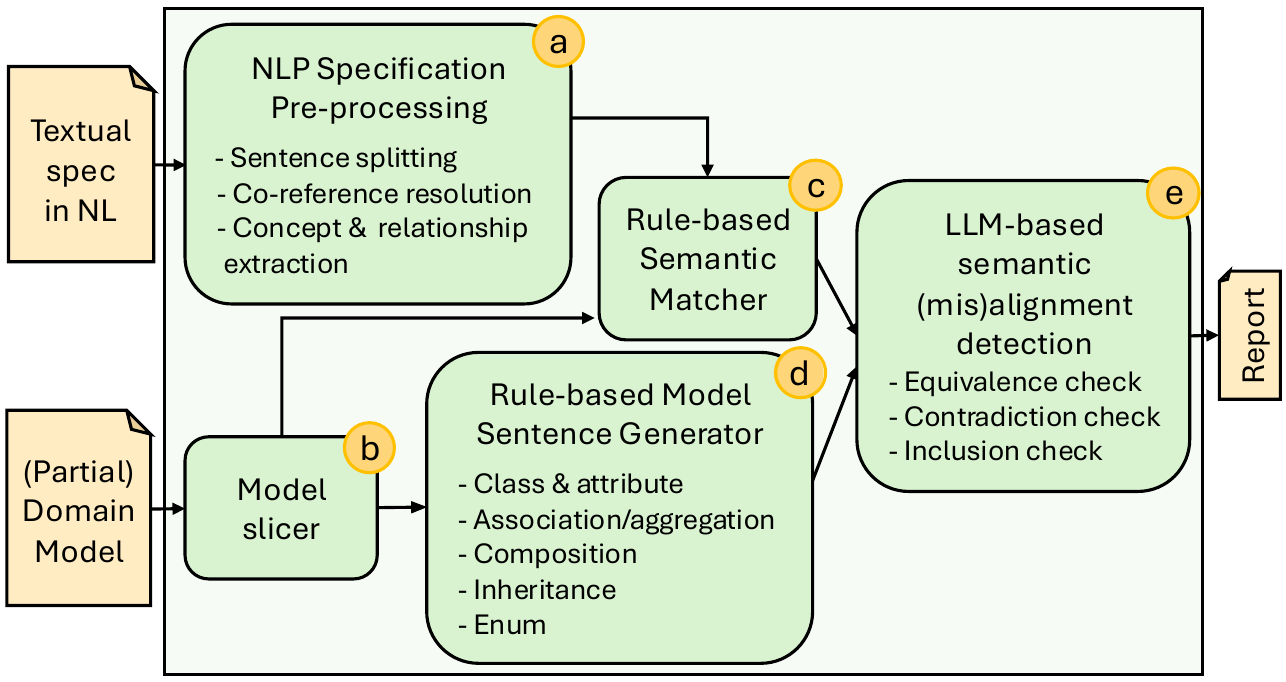}
    \caption{Approach Overview}
    \label{fig:overview}
\end{figure}

Figure~\ref{fig:overview} shows the overall structure of our approach. The system comprises five main components. Component A, the \textit{NLP Specification Preprocessor}, takes as input a requirements specification or problem description written in natural language. This module processes the description using rule-based natural language processing (NLP) to extract noun and relation tokens, which serve as candidates for possible text concepts ($tc$) and text relationships ($tr$), along with their traceability mappings.

Component B, the \textit{Model Slicer}, receives the domain model created by the modeler as input, which may still be under construction and thus represent a \textit{partial} domain model. This component traverses all attributes, associations, compositions, inheritance relationships, and enumerations in the domain model and extracts a minimal \textit{model slice} $m$ for each element.

The system passes the outputs of Component A (textual concepts and relationships) and Component B (model slices) to Component C. The Semantic Matcher aligns the textual concepts and relationships produced by the \textit{NLP Specification Preprocessor} with the model elements in the domain model. This process determines which sentences in the textual specification talk about which model elements, and it produces a set of matching specification sentences \{$sS$\} for each model element.

Component D, the \textit{Model Sentence Generator}, uses a rule-based algorithm to translate each model slice into a corresponding natural language sentence, providing an alternative description $mS$ for each model element.

The outputs of Component C and Component D are finally passed to Component E, \textit{Semantic Alignment Detection}, which uses LLMs to conduct three specific tests: the \textit{Semantic Equivalence} test, the \textit{Contradiction} test, and the \textit{Inclusion} test. These tests classify each model element as \textit{aligned}, \textit{misaligned}, or \textit{unclassified}. \insTwo{Note that instead of using more traditional approaches in components A, B, C and D we we could have used LLMs for all components in the pipeline. We preferred the more traditional approaches because 1) we had already created and validated components A and C in the past, and they outperformed the LLMs available at the time of running the experiment, 2) the implementation of components B and D was implemented very efficiently and deterministically in Python, and 3) using LLMs for all components would have incurred higher financial, energy and time costs.}

The following subsections now present the details of each component.

\subsection{NLP Specification Preprocessor}
\label{sub:nlp_preprocessor}
%In domain modeling, we transform informal requirements expressed in textual domain descriptions into domain models that represent analyzable and more precise
%specifications. These models represent domain concepts and their relationships in the form of classes, attributes, relationships, and cardinalities. As our approach aims to compare the domain models that result from domain description and manually constructed domain models, we need to first extract domain models from their textual domain descriptions.
To preprocess the textual specification, we use a simplified version of the analysis introduced in our previous work~\cite{saini2022automated}. In~\cite{saini2022automated}, we first use coreference resolution technique to identify all linguistic expressions that refer to the same entity or concept within a given specification, forming a \textit{references map}. This map facilitates pronoun resolution by replacing pronouns with their corresponding concept terms. Subsequently, the approach uses the \textit{spaCy} library~\cite{spacy} to process each statement sentence ($sS$) and construct a structure of linguistic features. This structure is further processed with rule-based NLP to extract \textit{noun chunks} and \textit{relation tokens} as candidates for possible \textit{text concepts} ($tC$) and \textit{text relationships} ($tR$), respectively. While the noun chunks are based on the linguistic features such as \textit{noun singular} (NS) part-of-speech (POS) tags, relation tokens are derived from the POS tags like \textit{verb} and \textit{preposition}. In this paper, we reuse and adapt this component to process each word in a noun chunk to determine the final set of text concepts using rule-based NLP. As an example, see the words highlighted in \textbf{bold} in the textual specification in Fig.~\ref{fig:running_example}. Next, the component identifies the sets of potential source and target text concepts %in addition to their cardinalities
for each token of relation and then uses additional heuristics to identify the final textual relationships. As an example, see \underline{underlined} words in the textual specification  in Fig.~\ref{fig:running_example}. %For example, the sentence 'Employee belongs to Department' needs additional rules to model the association relationship between 'Employee' and 'Department' classes based on the 'to' preposition.
The outputs of the component are 1) a set of tuples that map each text concept to the sentences that talk about the concept, and 2) a set of tuples that maps each relation to the source and target concept as well as the set of sentences that talk about the relation. Note that we only use modified sentences from the \textit{references map} for the extraction phase, but form the tuples using the original sentences \{sS\}. The output of the \textit{NLP Specification Preprocessor} component is formalised in equation~\ref{equ:output_component1}.
\begin{equation} 
\label{equ:output_component1}
\begin{aligned}
\text{TextualConcepts} &= \{(tC,\{sS\})\} \\ 
\text{TextualRelations} &= \{(tR, tC_{source}, tC_{target}, \{sS\})\}
\end{aligned}
\end{equation}

%\begin{equation} 
%\begin{aligned}
%M_e &= \{ C_1, C_2, \ldots, C_m, R_1, R_2, \ldots, R_n \} \\ 
%E_s &= \{ \text{ext\_sent}_1, \text{ext\_sent}_2, \ldots, \text{ext\_sent}_t \} \\ 
%T_e &= \{trace\_ele_1, trace\_ele_2, \ldots, trace\_ele_x \} \\ 
%\mathcal{M} &: T_e \to M_e \times \mathcal{P}(E_s)
%\end{aligned}
%\end{equation}

\ins{For example, for our running example, the \textit{NLP Specification Preprocessor} would output:}

\begin{equation} 
\label{equ:output_component1_running_example}
%\color{blue}
\begin{aligned}
\text{TextualConcepts} &= (car,\{s1,s2,s4\}), (service, \{s1,s3,s5,s6\}), \\ &\hspace{6mm} (plate, \{s2\}), (date, \{s3\}), ... \\ 
\text{TextualRelations} &= (happens, service, garage, \{s6\}), \\
&\hspace{5mm}(offers, garage, service, \{s1\}), ...
\end{aligned}
\end{equation}

\subsection{Model Slicer}
\label{subsec:slicer}
On the domain model side, the model slicer component traverses all attributes, associations, compositions, inheritance relationships and enumerations of the provided domain model and extracts a minimal \textit{model slice} for each of them. A minimal model slice, apart from the model element in focus, only contains other model elements that are necessary in order to obtain a valid model. For example, in the case of the domain model for our running example shown in Figure~\ref{fig:carMaintenanceModel}, the model slice extracted for the attribute \me{date} would include the attribute itself and also the class \me{Service}, because an attribute must be part of a class and the class sets the context of the attribute. Likewise, the model slice extracted for the association between the class \me{Service} and the class \me{Garage} would contain the association itself, the \me{Service} class and the \me{Garage} class, as well as the association ends containing the role name \me{place} with multiplicity \me{1} and the role name \me{provides} with multiplicity \me{*}.

The output of the \textit{Model Slicer} component is a set of tuples that map each model element of interest to a model slice as formalized in equation~\ref{equ:output_model_slicer}.

\begin{equation} 
\label{equ:output_model_slicer}
%\color{blue}
\begin{aligned}
\text{Slices} &= \{(m, slice)\}\text{, where } \\ 
slice &= \begin{cases}
  m\text{ is attribute: } class \cup attribute\\    
  m\text{ is association: } association \cup ends \cup both~classes\\
  m\text{ is inheritance: } inheritance \cup both~classes\\
  m\text{ is enum literal: } enumeration \cup enum~literal\\
  \text{otherwise: }m
\end{cases}
\end{aligned}
\end{equation}

\ins{For example, for our running example, the \textit{Model Slicer} would output:}
\begin{equation} 
%\color{blue}
\begin{aligned}
\text{Slices} &= (plate, \{Car, plate\}), (date, \{Service, date\}), ... \\
&\hspace{5mm}
(service\leftrightarrow
garage, \{service\leftrightarrow
garage, \\
&\hspace{5mm}place, provides, Service, Garage\}), ... 
\end{aligned}
\end{equation}

\subsection{Semantic Matcher}
\label{subsec:semantic_matcher}

The purpose of the \textit{Semantic Matcher} component is to relate the textual concepts and relationships produced by the \textit{NLP Specification Preprocessor} with the model elements in the domain model, with the ultimate goal of determining which sentences in the textual specification refer to which model elements.

To this aim, the matcher compares each model slice with the textual concepts and relationships from the \textit{NLP Specification Preprocessor} using a set of heuristics, i.e., syntactical word closeness and word similarity of the model element names and the textual concept names. While each slice focuses on one particular model element, all model elements of the slice are considered by the matcher. For example, for associations and inheritance, the connected classes are also used as additional information for the comparison. At this point, we also treat the association ends of associations and compositions separately, i.e., each association end (which includes role name and multiplicity) is considered a stand-alone model element. Association ends that do not have a multiplicity specified are not considered.

Whenever a match is made between a model element $m$ and a textual concept $tC$ or textual relation $tR$, the corresponding sentences from the textual specification are associated with the model element. In the end, a set of tuples associating each model element with a set of matched sentences of the textual specification is produced.

\begin{equation}
\begin{aligned}
\text{MatchedSentenceSets} = \{(m,\{sS\})\}
\end{aligned}
\end{equation}

\ins{For our running example, the \textit{Semantic Matcher} would output:}
\begin{equation}
%\color{blue}
\begin{aligned}
\text{MatchedSentenceSets} = (plate, \{s2\}), (place, \{s1,s6\}), ...
\end{aligned}
\end{equation}

%\begin{equation}
%\begin{aligned}
%T_f = \big\{ (m_i, gen\_sent_t, s_j) \mid\  
%& m_i \in \{C_m, R_n\}, 
%& s_j \in E_s, \\ 
%& \mathcal{M}(m_i) = s_j \big\}
%\end{aligned}
%\end{equation}

\subsection{Sentence Generator}
\label{subsec:sentence_generator}

% \lola{if an association end does not have a multiplicity, we do not generate a sentence bc we consider it is underconstruction}

%The model element sentence generator first splits a (partial) domain model into different groups of model elements where each group represents a unique \textit{model slice}. We consider a \textit{model slice} at a granularity level where we can construct atomic sentences for the underlying group of model elements. For example, the Service class with two attributes~\ref{fig:carMaintenanceModel} represents two \textit{model slices} (Service class with date attribute and Service class with type attribute). We can represent the set of model slices \( M_s \), which consists of each \textit{model slice} \(model\_sc_t \) for a given (partial) domain model, as follows:
%\begin{equation}
%\begin{aligned}
%M_s &= \{ model\_sc_1, model\_sc_2, \ldots, model\_sc_t \}
%\end{aligned}
%\end{equation}

The \textit{Sentence Generator} component constructs sentences in natural language for each model element of interest in the domain model from the corresponding \textit{model slice} using rule-based NLP. We based our rules on the rules proposed by Arora et al.~\cite{10.1145/2976767.2976769} and extended them further to support additional model elements such as enumerations and association role names. The final output of this component is a set of tuples that maps each model element to a generated sentence:
\begin{equation}
\begin{aligned}
\text{GeneratedSentences} &= \{ (m, mS) \}
\end{aligned}
\end{equation}

In the remainder of this subsection, we further explain the rules the \textit{Sentence Generator} uses to build sentences for attributes, associations, compositions, inheritance, and enumerations.

\subsubsection{Attributes}
We process the word(s) in an attribute name using the \textit{spaCy} library to obtain linguistic features such as nouns and verbs. The text generation algorithm employs different rules to generate a sentence based on these features. For example, the generated sentence for the model slice focusing on the \me{plate} attribute in the class \me{Car} in Figure~~\ref{fig:carMaintenanceModel} results in the sentence `A car has a plate.'.

\subsubsection{Association (and Aggregation) Relationships}
Since aggregation is a special type of association relationship whose semantics are not fully defined in the UML~\cite{omg2015umls}, our approach treats aggregation in the same way as plain associations.
%While constructing a textual representation of an association relationship, we generate two sentences, one for each end of the relationship. Only one sentence is generated for an aggregation relationship, as modelers usually indicate the role name and multiplicity only on one end.
In our approach, one sentence is generated for each association end for which a multiplicity is provided.
First, we tokenize the role name using \textit{spaCy} to extract linguistic features such as nouns and verbs. Second, we derive a sentence using the role names. If the role name is absent or the role name does not contain a verb, we use the auxiliary verb `has' (`have' in case of a plural) to complete the sentence. For example, the \textit{model slice} consisting of an association relationship from the \me{Service} class to the \me{Garage} class results in the two sentences `A service has a place which is a garage.' and `A garage provides services.'.

\subsubsection{Composition Relationships}
For compositions, we use the wording `is made up of' to construct the sentence. For example, the \textit{model slice} consisting of the composition relationship between the \me{Car} and \me{Part} classes leads to the sentence `A car is made up of parts.'.

\subsubsection{Inheritance Relationships}
For inheritance, we generate one sentence for the subclass that is present in the \textit{model slice}. We use the wording `is a type of'. For example, the \textit{model slice} with the \me{Service} and \me{Repair} classes in Figure~\ref{fig:carMaintenanceModel} results in the sentence `Repair is a type of service.'. %Similarily, for the \textit{model slice} with the Service and Maintenance classes~\ref{fig:carMaintenanceModel} results in the sentence 'Maintenance is a type of a service'.

\subsubsection{Enumerations}
For enumerations, we generate one sentence for each enumeration literal. We use the wording `is a/an' to generate the sentence. For example, the model slice that has an enumeration \me{PartType} with \me{ENGINE} as an enumeration literal, leads to the sentence `Engine is a part type.'. %Similarly, we construct sentences for the \textit{model slice} with 'Transmission' and 'Lights' as enumeration literals.

\vspace{2mm}

\ins{Based on the above explanations, for our running example, the generated sentences would look as follows:
}
\begin{equation}
%\color{blue}
\begin{aligned}
\text{GeneratedSentences} &= (plate,\text{``A car has a plate.''}), \\
&\hspace{5mm}(place,\text{``A service has a place.''}), \\
&\hspace{5mm}(provides, \text{``A garage provides services.''}), ...
\end{aligned}
\end{equation}

\subsection{LLM-based Semantic (Mis)Alignment Detection}

\begin{figure}[ht]
\centering
    \includegraphics[width=0.75\textwidth]{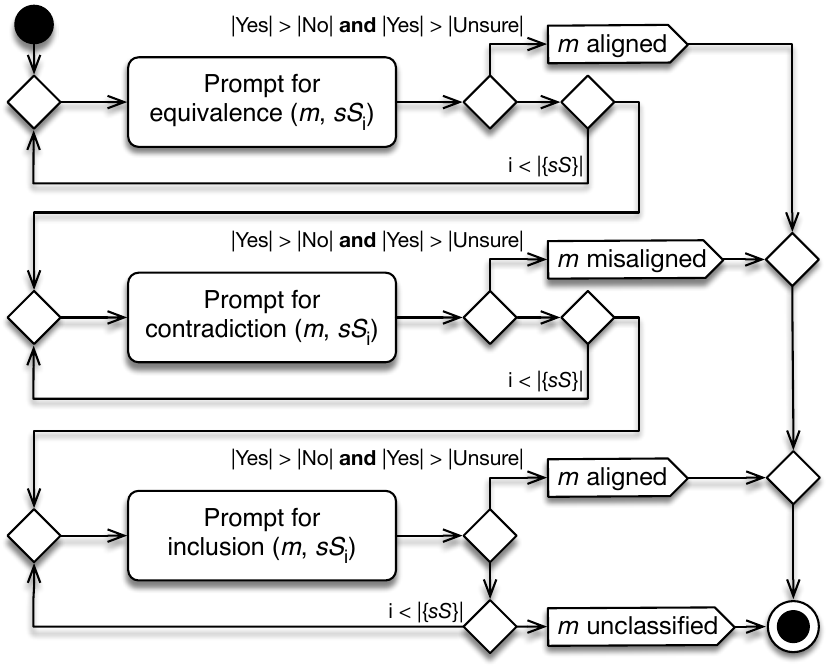}
    \caption{Classification Workflow}
    \label{fig:workflow}
\end{figure}

The most advanced component of our approach is the \textit{LLM-based Semantic (Mis)Alignment Detection}\footnote{We refer to alignments and misalignments collectively as (mis)alignments.} component (component E in Figure~\ref{fig:overview}). It queries an LLM using the combined information produced by the \textit{Semantic Matcher} and the \textit{Sentence Generator} to decide, for each model element, whether it is \textit{aligned} with the textual specification (i.e., \textit{correct}), \textit{misaligned} with the textual specification (i.e., \textit{incorrect}), or whether it leaves it \textit{unclassified} (i.e., there is not enough evidence to make a decision). It does so by asking the LLM three sets of prompts as shown in the overview Figure~\ref{fig:workflow}. The detailed algorithm is described in subsection~\ref{subsubsec:algorithm}.

\subsubsection{Prompt Engineering}
\label{subsubsec:prompt_engineering}
We performed some initial experiments with different LLMs (GPT 3.5, GPT 4 and Gemini) using both zero shot and few shot prompting and empirically determined that: 1) GPT 4 performed the best; 2) few shot prompting did not improve the performance; 3) the responses of the LLM vary considerably depending on the complexity of the statements, the domain, and the exact prompt wording.
Due to the first two points, we decided to use GPT~4 with zero shot prompting. 

To illustrate the last point, consider the following example, where we need to determine whether the original statement ``A football player can be sold to another football team.'' includes the generated statement ``A football player plays for a football team.''. The LLM's answer to the question ``Can the generated statement be \textit{implied} from the original statement?'' was ``Yes, the generated statement can be implied from the original statement. The fact that a football player can be sold to another football team indicates that they play for a football team in the first place.'', whereas the LLM's answer to the question ``Can the generated statement be \textit{derived} from the original statement?'' was ``No, the generated statement cannot be derived from the original statement. While both involve the concept of a football player being part of a football team, the original statement doesn't inherently confirm the fact stated in the generated statement.''.

To deal with these cases with variance in responses, we decided to ask the LLM \textit{multiple} diverse, but semantically equivalent questions and use relative majority voting to determine the final result. This technique is inspired by voting techniques used in fault tolerance, e.g., N-version programming~\cite{chen1995}.

To distill a set of diverse, yet semantically equivalent prompts for testing \textit{Equivalence}, \textit{Contradiction}, and \textit{Inclusion} we ran a meticulous experimentation and analysis on five representative domain models consisting of a total of 26 model elements. For each type, we started with a larger number of prompts, asking the LLM to answer \textit{Yes}, \textit{No} or \textit{Not Sure}. We then selected the final prompts by eliminating those that provided false positives (i.e., incorrect answers), as well as those that did not introduce any diversity in the responses, i.e., provided exactly the same answers. Finally, since the \textit{Semantic Matcher} component at times determines matches with the sentences from the problem description that are actually unrelated, we also gave preference to the prompts that for unrelated sentences correctly output \textit{Unsure}, and made sure that in the end we have an odd number of prompts.

For example, for the \textit{Equivalence} test, we ended up with the following five prompts:

\vspace{1mm}
{\small
\noindent
\fbox{%
\parbox{0.97\columnwidth}{%
\textbf{Equivalence Prompt 1}:\\
Are the following two statements semantically equivalent?\\
Statement 1: \color{gray}A car has a plate.\\
\color{black}Statement 2: \color{gray}For each car that comes to the garage, the first thing to do is to register its
plate number.\\ \color{black}
\textbf{Equivalence Prompt 2}:\\
Do the following two statements convey the same information? [...]\\
\textbf{Equivalence Prompt 3}:\\
Are the following two statements conveying the same meaning? [...]\\
\textbf{Equivalence Prompt 4}:\\
Are these statements synonymous? [...]\\
\textbf{Equivalence Prompt 5}:\\
Do statement 1 and statement 2 have identical implications? [...]
}
}
}\\

\ins{Checking for contradiction is done in a similar way as checking for equivalence, but using the following questions in the prompt: `Do these statements contradict each other?', `Are these statements mutually exclusive?', `Do these statements clash or conflict with each other?', `Do these statements negate each other?', `Are these statements inconsistent?', `Are these statements in disagreement?', and `Are these statements incompatible?'.}

\ins{Finally, for checking inclusion, the questions in the prompts sent to the LLM are: `Can Statement 1 be inferred from Statement 2?', `Can Statement 1 be implied from Statement 2?', `Can Statement 1 be determined from Statement 2?', `Can Statement 1 be derived from Statement 2?', `Does Statement 1 logically follow from Statement 2?', `Can Statement 1 be concluded based on Statement 2?' and `Does Statement 2 support Statement 1?'.}

\subsubsection{Algorithm}
\label{subsubsec:algorithm}

The pseudocode of the algorithm that performs the sentence comparisons to classify the model elements is described in Algorithm~\ref{alg:algoritm}.

\begin{algorithm}
\vspace{-2.5cm}
\scriptsize

\DontPrintSemicolon
\textbf{Input:} \\
\hspace{1mm} model: \dotfill $\{m\}$\\
\hspace{1mm} generatedSentences (one tuple for each $m$): \dotfill $\{ (m,mS) \}$\\
\hspace{1mm} matchedSentenceSets (one tuple for each $m$): \dotfill \{($m,\{sS\})\}$\\

\textbf{Output:}\\
\hspace{1mm} aligned $m$ and sentences: \dotfill $\{(m_{aligned},\{sS_{aligned}\})\}$, \\
\hspace{1mm} misaligned $m$ and sentences: \dotfill  $\{(m_{misaligned},\{sS_{misaligned}\})\}$, \\
\hspace{1mm} unclassified $m$: \dotfill $\{m_{unclassified}\}$ 
\\
\vspace{2mm}
\textbf{def} classify(model, generatedSentences, matchedSentenceSets)
\Begin{
    \textit{alignments} $\leftarrow$ $\emptyset$\; \label{line:alg1_begin_init}
    \textit{misalignments} $\leftarrow$ $\emptyset$\;
    \textit{unclassified} $\leftarrow$ $\emptyset$\; \label{line:alg1_end_init}
    \For {$m$ \textbf{in} model} {
        \label{line:alg1_begin_for} \textit{aligned} $\leftarrow (m, \emptyset)$\;
        \textit{misaligned} $\leftarrow (m,\emptyset)$\;
        \textit{included} $\leftarrow (m,\emptyset)$\;
        \For {$sS$ \textbf{in} matchedSentenceSets($m$)} {
            \label{line:alg1_begin_inner_for} \textit{mType} $\leftarrow$ getType($m$)\;
            \eIf {checkForEquivalence \label{line:alg1_call_eq}\\
            \hspace{5mm}($sS$, generatedSentences($m$), mType)} {
                \textit{aligned}.addSentence($sS$)
            } {
                \eIf {\textit{aligned}.containsNoSentences \textbf{and} checkForContradiction \label{line:alg1_call_cont}\\
                \hspace{5mm}($sS$, generatedSentences($m$), mType)} {
                    \textit{misaligned}.addSentence($sS$)
                } {
                    \If {checkForInclusion \label{line:alg1_call_incl}\\
                    \hspace{5mm}($sS$, generatedSentence($m$), mType)} {
                        \textit{included}.addSentence($sS$)
                    }
                }
            }
        } \label{line:alg1_end_inner_for} 
        \eIf {\textit{aligned}.containsSentences() \label{line:alg1_begin_decision}} {
            \textit{alignments}.addTuple(\textit{aligned})
        }
        { \eIf {\textit{misaligned}.containsSentences()} {
            \textit{misalignments}.addTuple(\textit{misaligned})
        }
        { \eIf {\textit{included}.containsSentences()} {
            \textit{alignments}.addTuple(\textit{included})
        }
        {
            \textit{unclassified}.add($m$)
        } } } \label{line:alg1_end_decision}
    } \label{line:alg1_end_for}
    \nl\KwRet{$alignments$, $misalignments$, $unclassified$}
}
\vspace{2mm}
\textbf{def} checkForEquivalence($sS$, $mS$, $mType$) \label{line:alg1_begin_eq}
\Begin{
    \textit{prompts} $\leftarrow$ createEquivalencePrompts($mType$, $mS$, $sS$)\;
    \textit{yes} $\leftarrow$ 0, \textit{no} $\leftarrow$ 0, \textit{unsure} $\leftarrow$ 0\;
    \For {pt \textbf{in} prompts} {
        \textit{response} $\leftarrow$ prompt(llm, pt)\;
        \Switch{parse(response)} {
        \textbf{case} "Yes": \textit{yes} = \textit{yes} + 1\;
        \textbf{case} "No": \textit{no} = \textit{no} + 1\;
        \textbf{case} "Unsure": \textit{unsure} = \textit{unsure} + 1\;
        }
    }
    \KwRet{$yes > no$ \textbf{and} $yes > unsure$} \label{line:alg1_majority}
} \label{line:alg1_end_eq}

// Similar code for the contradiction and inclusion checks\\
\caption{LLM-based (Mis)Alignment Detection}

\label{alg:algoritm}
%\lola{the for is not at a tuple level but at a modeling element level.
%For each model element, we see what the sentences from the textual specification that match and we study eq, contr and inclu.
%if eq = 1, cont = 2, nonclass=5
%Once we have the result of all pairs, if there is any contradiction, then we mark it as contradiction. otherwise if there is any equivalence $->$ eq, otherwise non classified.}
\end{algorithm}

% %{\small
% \noindent\fbox{%
%     \parbox{\columnwidth}{%
%     \textbf{Specification:}
%     \textcolor{red}{$^1$}Cities put in place bike rental systems to encourage their citizens to be more sustainable. 
%     \textcolor{red}{$^2$}For each city, we need to store its name and country.
%     \textcolor{red}{$^3$}Cities have bike stations located at different addresses.
%     \textcolor{red}{$^4$}Apart from the address, for each bike station we need to store its name and the number of spots that it has to park bikes.
%     \textcolor{red}{$^5$}When bikes are not in use, they are parked in bike stations. For each bike, we store its code as an integer and the price per hour.Note that different bikes might have different prices per hour. 
%     \textcolor{red}{$^6$}Bikes are rented by users.
%     \textcolor{red}{$^7$}For each user we need to keep their id, name and credit card information.
%     \textcolor{red}{$^8$}For each rental, we need to keep track of the start and end date, the pickup station, the drop-off station, the user who has rented the bike and the bike that has been rented.
%     }%
% }
% %}

After initializing the output data structures (lines~\ref{line:alg1_begin_init}-\ref{line:alg1_end_init}), the main \texttt{for} loop iterates through all model elements in the domain model. For each model element $m$, a nested \texttt{for} loop goes through all matched sentences from the textual specification (previously identified by the \textit{Semantic Matcher}). For each matched sentence $sS$, the algorithm asks the LLM whether $sS$ is equivalent to $mS$, the generated sentence for $m$ (produced by the \textit{Sentence Generator}) by calling the function \texttt{checkForEquivalence} on line~\ref{line:alg1_call_eq}. 

How this is done is explained in detail in the pseudocode of the function \texttt{checkForEquivalence} on lines~\ref{line:alg1_begin_eq}-\ref{line:alg1_end_eq}.
First, an \textit{odd number} of semantically equivalent prompts are created using the five diverse questions listed in the previous subsection. 
Then, to determine whether the statements are considered equivalent or not, we use a relative majority vote (see line~\ref{line:alg1_majority}).
If the two sentences are not judged to be equivalent, the algorithm proceeds to ask the LLM whether the two sentences contradict each other by calling \texttt{checkForContradiction} on line~\ref{line:alg1_call_cont}. \del{Checking for contradiction is done in a similar way as checking for equivalence, but using the following questions in the prompt: `Do these statements contradict each other?', `Are these statements mutually exclusive?', `Do these statements clash or conflict with each other?', `Are these statements at odds with each other?', `Do these statements negate each other?', `Are these statements inconsistent?', `Are these statements in disagreement?', `Are these statements incompatible?' and 'Do these statements present conflicting viewpoints?'.}
If the two sentences are also not judged to be contradicting, then the algorithm finally asks the LLM whether the meaning of the generated sentence is included in the meaning of the sentence from the textual specification by calling \texttt{checkForInclusion} on line~\ref{line:alg1_call_incl}. \del{This time, the questions in the prompts sent to the LLM are: `Can Statement 1 be inferred from Statement 2?', `Can Statement 1 be implied from Statement 2?', `Can Statement 1 be determined from Statement 2?', `Can Statement 1 be derived from Statement 2?', `Does Statement 1 logically follow from Statement 2?', `Can Statement 1 be concluded based on Statement 2?' and `Does Statement 2 support Statement 1?'.}

Once all the matched sentences for the model element $m$ have been compared with the generated sentence and classified as \textit{aligned}, \textit{misaligned} or \textit{included}, the final decision for $m$ is obtained as shown in the pseudocode in lines~\ref{line:alg1_begin_decision}-\ref{line:alg1_end_decision}. If at least one of the matched sentences was judged to be equivalent, $m$ is classified as \textit{aligned}. Only if no sentence was judged equivalent, then, in the case where a contradiction was detected, the model element is judged as \textit{misaligned}. Finally, if no contradiction was detected, but at least one of the matched sentences was judged to include the generated sentence, then $m$ is judged to be \textit{aligned}.

The rationale for doing it this way is the following: All the LLMs that we tested our approach with were quite critical when being asked about equivalence. If the sentence from the specification provided only a little bit more detail, or talked about the model element in the context of an action\footnote{This can happen often in textual specifications that are based on user stories or use cases.}, then the LLM would not consider the sentence equivalent to the generated one. Because of that, our algorithm gives equivalence the highest priority.

Contradiction has second highest priority, which can again be explained by the fact that for the LLM to consider two sentences to contradict they really have to portray the same properties in the same context in an inconsistent way.

Finally, when the LLM judges the generated sentence to be included in one or several of the matched sentences from the textual specification, then we also consider the model element to be aligned. This is because sentences in textual specifications often describe multiple properties at once, while these properties are modeled as separate model elements in the domain model. For example, when the textual specification sentence is a composite sentence containing a conjunction, e.g., `and', then very likely the textual specification contains more information than the generated sentence, and hence would be judged as not equivalent, but as \textit{including} the generated sentence. This is the case in the sentence `For each service provided, we record the date
and the type of service.' of our running example. This \textit{one} sentence is related to \textit{two} attributes, namely the attributes \me{date} and \me{type} of the class \me{Service} in Figure~\ref{fig:carMaintenanceModel}. Hence, the sentence is not equivalent to, but includes the generated sentences `A service has a date.' and `A service has a type.'.

\subsection{Tool Support}
\label{subsec:tool_support}
We have developed a proof of concept implementation for our approach \ins{and made it available in a Git repository~\cite{repo}. The UML class diagrams are provided in \texttt{.cdm} format, an XML-based format used by the TouchCORE modeling tool~\cite{schiedermeier_multi-language_2021}, which itself is implemented using the Eclipse Modeling Framework. The implementation of the components of our approach, including the extraction of the model elements from the \texttt{.cdm} files, as well as the automation of the data processing shown in Figure~\ref{fig:overview} is done in Python. The Python script} uses NLP libraries such as \emph{spaCy} for NLP preprocessing, \emph{Stanza} for co-reference resolution, \emph{Inflect} for operations related to string, \emph{NumPy} for numerical operations, and {pandas} to store results and CSV file operations. As LLM we have chosen GPT-4o due to its advanced reasoning capabilities. We used the following hyperparameters: temperature=1, top\_p=1, max\_completion\_tokens=2048, frequency\_penalty=0, presence\_penalty=0. 

\del{This implementation is available in our Git repository~\cite{repo}.}

% This proof of concept is available in our anonymous repository~\cite{repo}.
% \lola{Shwetali, maybe we need to describe a little bit the implementation.} 
% Librraies used : Spacy for NLP preprocessing, en-core-web-trf model of spacy,
% Stanza(Coreference resolution), 
% Inflect for operations related to string
% numpy(for numerical operations), 
% pandas(to store results and csv file operations)

Let us note that the proposed approach remains agnostic to the programming language, the selected libraries and LLM, as these could be replaced by similar or improved technologies.

\section{Validation}
\label{sec:evaluation}

In the next subsections, we describe our research questions, our empirical setup including the textual specifications and models that we have used for validation and the metrics we have used to answer the research questions. We present the results of our evaluation and discuss the answer to each research question. Finally, we discuss the threats to validity.

For transparency and reproducibility purposes, the dataset, scripts to reproduce our results and excel sheets with all the logs and results of our experiments are available in our repository~\cite{repo}.

\subsection{Research Questions}

To evaluate our approach, we have defined three research questions regarding the correctness (exactness) of the obtained (mis)alignments, the completeness (coverage) of the identified (mis)alignments, and the scalability of our approach.

% Our study addresses the following three research questions. With these questions, we aim to justify the usefulness of our approach.

\begin{questions}
\item Correctness. Of all the alignments and misalignments identified by our approach, how many are correct?

\item Completeness. Of all the existing alignments and misalignments, how many are correctly identified by our approach?

%\emph{RQ3. Execution performance. How long does our approach take to detect alignments and misalignments?} 
\item Scalability. How well does our approach scale as the size of the textual specification and corresponding domain model increase?
\end{questions}

\subsection{Experimental Setup}

% In this work, we have developed a proof of concept for our approach. Its implementation demonstrates the feasibility of our approach as serves us for validation. 

% The proof of concept has been implemented in Python. As LLM, we have used GPT-4o by OpenAI. The hyperparameters used are a \texttt{temperature} of \todo{1} and a value of \todo{4096} for \texttt{maxTokens}.
% Parameter values with GPT:
% model="gpt-4o",
%         temperature=1,
%         top_p=1,
%         max_completion_tokens=2048,
%         frequency_penalty=0,
%         presence_penalty=0,

% This proof of concept is available in our anonymous repository~\cite{repo}.
% \lola{Shwetali, maybe we need to describe a little bit the implementation.} 
% Librraies used : Spacy for NLP preprocessing, en-core-web-trf model of spacy,
% Stanza(Coreference resolution), 
% Inflect for operations related to string
% numpy(for numerical operations), 
% pandas(to store results and csv file operations)

% \lola{Briefly report on the technical setup of the experiments, the tools we developed, which LLMs we used and why}

\subsubsection{Dataset}

\begin{table*}[t]
    \centering
    \caption{Dataset Characterization}
    \label{tab:dataset}
    \scriptsize
    \begin{tabular}{c|c|c|c|c|c|c|c|c|c}

  &Domain   & words & sentences & classes & attr. & \begin{tabular}[c]{@{}c@{}}assoc./\\ aggr.\end{tabular} & comp. & enums & inh. \\ \hline
  R1 & Restaurant Management& 255  & 20 & 10 & 8 & 12&1 &2 & 1\\
  R2 &Employee Management& 227 & 19 & 8 & 13 & 3 & 0 & 1 & 5\\ 
  R3 & Library & 209 & 22 & 7 & 13 & 4 & 0 & 2 & 2 \\
  R4 & Galaxy Sleuth Game & 710 & 39 & 10 & 12 & 14 & 2 & 0 & 0 \\
  R5 &Spy-Robot Game& 412 & 17 & 8 & 5  & 3  & 0  & 3 & 2  \\
  R6 &Academic Program& 129 & 9 & 8 & 13 & 4 & 0 & 0 & 0 \\
  R7 &Supermarket& 543 & 21 & 9 & 4 & 6 & 3 & 2 & 1 \\
  R8 &Hotel Reservation& 312 & 20 & 7 & 5 & 2 & 0 & 1 & 3 \\
  R9 &BeWell app& 433 & 26 & 12 & 5 & 7 & 4 & 0 & 1 \\
  R10 &File Manager& 83 & 7 & 6 & 6 & 1 & 1 & 0 & 3 \\
  R11 &Football team& 135 & 12 & 3 & 4 & 1 & 0 & 1 & 0 \\
  R12 &Rented Car Gallery& 120 & 12 &3  &9  &3  &0  &0  &0  \\
  R13 &Course Enrollment& 234 & 15 & 7 & 6 & 6 & 2 & 0 & 0 \\
  R14 &ATM& 164 & 10 & 9 & 0 & 5 & 3 & 0 & 0 \\
  R15 &Video Rental& 221 & 17 & 6 & 5 & 4 & 1 & 0 & 0 \\
  R16 &Cinema& 172 & 9 & 4  & 4  & 3  & 0  & 0  & 0  \\
  R17 &Timbered House& 95 & 6 & 6 & 4 & 6 & 0 & 0 & 0 \\
  R18 &Musical store& 147  & 17  & 4 &5  &5  &0  &1  &1  \\
  R19 &Airport& 303 & 20  &11  &7  &15  &1  &0  &0  \\
  R20 &Monitoring Pressure& 92  & 6  & 2 & 2  &1  &0  &0  &0  \\
  R21 &Savings Account& 381 & 24 & 2  &7  &1  &0  &2  &0  \\
  R22 &IPO application& 411 & 29 & 3 & 13 & 2 & 0 & 1 & 0 \\
  R23 &PIN& 422 & 26 &2  &4  &1  &0  &1  &0  \\
  R24 &Communication Prefs& 311  & 17  & 1  &3  &0  &0  &1  &0  \\
  R25 &Apple Pay& 301 & 17  &3  &1  &2  &0  &2  &1  \\
  R26 &Block Card& 290  & 19  &3  &3  &2  &0  &3  &1  \\
  R27 &Biometric Login& 493 & 30 &5  &7  &2  &0  &1  &2  \\
  R28 &Donation& 361  & 25  &6  &3  &5  &0  &0  &1  \\
  R29 &Prepaid Card& 339  & 20  &4  &4  &2  &0  &2  &1  \\
  R30 &Wallet app& 251  & 17  &4  &3  &2  &1  &1  &0  \\ \hline
  \begin{tabular}[c]{@{}c@{}}Avg. \\ $\pm$ Std. \end{tabular} & All & 
  \begin{tabular}[c]{@{}c@{}}285.2\\ $\pm$ 148.0 \end{tabular}
  &
  \begin{tabular}[c]{@{}c@{}}18.3\\ $\pm$ 7.6 \end{tabular}
  &
  \begin{tabular}[c]{@{}c@{}}5.8\\ $\pm$ 3.0 \end{tabular}
  & 
  \begin{tabular}[c]{@{}c@{}}5.9\\ $\pm$ 3.7 \end{tabular}
  & 
  \begin{tabular}[c]{@{}c@{}}4.1\\ $\pm$ 3.7 \end{tabular}
  & 
  \begin{tabular}[c]{@{}c@{}}0.6\\ $\pm$ 1.1 \end{tabular}
  & 
  \begin{tabular}[c]{@{}c@{}}0.9\\ $\pm$ 1.0 \end{tabular}
  & 
  \begin{tabular}[c]{@{}c@{}}0.8\\ $\pm$ 1.2 \end{tabular}
  \\
    \end{tabular}
\end{table*}

To evaluate our approach, we chose a publicly available dataset of 120 textual software requirements written in English~\cite{chaudron2024}. According to the authors, these requirements have been acquired from industry (48\%) as well as academia (52\%), and carefully selected to cover a wide range of domains (business, finance, health, entertainment, education, technology, energy) and complexity. In the same paper, the authors use the first 30 requirements to compare two domain model extraction tools, and they published the domain models created from those 30 requirements. These textual requirements as well as the models are available on the IEEEDataPort\footnote{\url{https://ieee-dataport.org/documents/dataset-text-requirements- models}.}.

For our validation, we have chosen to use the same 30 textual requirements as the authors of~\cite{chaudron2024} used for their experiment, as well as the domain models they provided (when possible). We observed that the domain models they published were either manually created or ChatGPT-generated. Furthermore, some domain models were expressed by means of a class diagram, while others were expressed in structured text.

In order to have a homogeneous dataset for our evaluation consisting of pairs of textual requirements and the corresponding domain model created by a human expert and expressed as a class diagram, for those descriptions whose domain model was generated by ChatGPT, we discarded the models from by~\cite{chaudron2024} and asked a professor (who teaches a modeling course) to create the domain model. We also asked this professor to convert the structured text into a domain model expressed using a class diagram.
All the textual requirements and corresponding domain models are available in our repository~\cite{repo}.
% The textual requirements by the authors of~\cite{chaudron2024} are available on the IEEEDataPort\footnote{\url{https://ieee-dataport.org/documents/dataset-text-requirements- models}.}, as well as the domain models they provided for the first 30 requirements in structured text format. 

%, and the available domain models after transforming them from structured text to UML.

We used these models as \textit{ground truth}, i.e., all elements of the model are considered correct~/~aligned with the textual specification. The resulting pairs of 30 textual requirements and domain models are listed in Table~\ref{tab:dataset}, together with information about the number of words and sentences in the text, as well as the number of classes, attributes, associations, compositions, enumerations and inheritance relationships in the domain models.

Since we also want to be able to evaluate the misalignment detection performance of our algorithm, we also needed domain models that contain errors. 
According to the IEEE Standard Classification for Software Anomalies~\cite{ieee_standard_anomalies}, defects in artifacts such as models or code are due to elements that are either \textit{missing}, \textit{wrong} (i.e., inconsistent, incorrect or ambiguous), or \text{unnecessary} (redundant or extraneous). Since our approach aims to detect misalignments of \textit{existing model elements} and should also work on partial models, we do not focus on detecting defects resulting from missing elements. Likewise, since during the process of domain modeling, experienced modelers often discover additional model elements that are not explicitly found in the problem description, we are also not concerned with detecting unnecessary model elements. In other words, our approach for misalignment detection focuses on finding \textit{wrong} model elements, in particular \textit{inconsistent} and \textit{incorrect} ones.

In~\cite{granda2015}, the authors present a systematic mapping study that identifies 226 different kind of defects that have been detected in 28 primary studies involving model-driven development and classified them. In~\cite{granda2016}, the same authors focused on defects found in domain models, and defined a set of 50 mutation operators that are applicable to domain models to produce the kind of defects encountered in~\cite{granda2015}. From this list of 50, many are related to adding or removing model elements, i.e., they introduce errors due to missing or unnecessary information. After inspection, only the following three mutation operators from~\cite{granda2016} are applicable to our specific context:

\begin{itemize}
\item 29-WAS$2^*$: Change the type of an association type, i.e., from normal to composite;
\item 30-WAS$4^*$: Change the multiplicity of an association member end, i.e., from 0..1 to 0..* or from 1..1 to 1..*, or vice-versa;
\item 31-WGE: Change the member ends of a generalization, i.e., change the superclass or the subclass.
\end{itemize}

To introduce errors into the models in an unbiased way, while at the same time ensuring that we do not neglect a mutation operator, we applied the following strategy. For each model in our dataset and for each mutation operator, we determined the number of model elements in the model that the operator can be applied to. We then randomly applied the mutation operator to 20\% of those model elements (rounding up).
Due to the fact that the use of a mutation operator could still potentially produce a correct model with respect to the specification (especially when the specification is vague), we have manually checked that the mutated element actually introduced an error, i.e., it deviates from the information provided in the specification.

\subsubsection{Evaluation Metrics}

To answer our research questions, we use several metrics depending on the nature of the question.

To evaluate RQ1 (correctness), we use the precision when calculating alignments (Precision$_a$), when calculating misalignments (Precision$_m$) and considering both alignments and misalignments at the same time (Precision).
These precisions are defined as follows:

{\small
\noindent
\begin{tabularx}{\linewidth}{@{}>{\centering\arraybackslash}X>{\centering\arraybackslash}X@{}}
\begin{equation}
  \text{Precision$_a$} = \frac{|\text{CPA}|}{|\text{PA}|}
\end{equation}
&
\begin{equation}
  \text{Precision$_m$} = \frac{|\text{CPM}|}{|\text{PM}|}
\end{equation}
\end{tabularx}

\begin{equation}
\text{Precision} = \frac{|\text{CPA}| + |\text{CPM}|}{|\text{PA}|+ |\text{PM}|}
\end{equation}
}
where
CPA = correctly predicted alignments, PA = predicted alignments, CPM = correctly predicted misalignments, and PM = predicted misalignments.

%{\small
%$$ \text{Precision$_a$} = \frac{|\text{alignments predicted and actually aligned}|}{|\text{alignments predicted}|} $$
%$$ \text{Precision$_m$} = \frac{|\text{misalignments predicted and actually misaligned}|}{|\text{misalignments predicted}|} $$
%$$ \text{Precision$_{am}$} = \frac{|\text{(mis)alignments predicted and actually (mis)aligned}|}{|\text{(mis)alignments predicted}|} $$
%}

RQ2 (completeness) can be answered by calculating the recall. Similarly, as we have done for RQ1, we calculate the recall separately for alignments and misalignments, as well as combined. For this, we use the following formulas:

{\small
\noindent
\begin{tabularx}{\linewidth}{@{}XX@{}}
\begin{equation}
  \text{Recall$_a$} = \frac{|\text{CPA}|}{|\text{A}|}
\end{equation}
&
\begin{equation}
  \text{Recall$_m$} = \frac{|\text{CPM}|}{|\text{M}|}
\end{equation}
\end{tabularx}

\begin{equation}
    \text{Recall} = \frac{|\text{CPA}| + |\text{CPM}|}{|\text{A}| + |\text{M}|}
\end{equation} 
}

where A = alignments (according to the ground truth), and M = misalignments (according to the ground truth). 

%{\small
%$$ \text{Recall$_a$} = \frac{|\text{alignments predicted and actually aligned}|}{|\text{actual alignments}|} $$
%$$ \text{Recall$_m$} = \frac{\text{misalignments predicted and actually misaligned}}{|\text{actual misalignments}|} $$
%$$ \text{Recall$_{am}$} = \frac{|\text{(mis)alignments predicted and actually (mis)aligned}|}{|\text{actual (mis)alignments}|} $$
%}

To complement the quantitative values obtained for RQ1 and RQ2, we also report on the F$_1$-score. The F$_1$-score is the harmonic mean of precision and recall:

{\small
\begin{equation}
\text{F}_1\text{-score} = 2 \times \frac{\text{Precision} \times \text{Recall}}{\text{Precision} + \text{Recall}}
\end{equation}
}

Finally, to study the scalability of our approach (RQ3), we need to take into account two main factors: the size of the textual requirements in number of words ($w$) and the number of model elements in the domain model ($m$). We will elaborate the complexity of each component in our approach with respect to those two variables. Furthermore, to give an indication of the performance of our approach, we provide the \ins{time per component and the} total time to run our approach on a subset of the models.
%Therefore, to answer this question we calculate different values. For each example in our dataset, we calculate the total time needed by our approach, the total time needed by each component, and the time needed to calculate each (mis)alignment. We provide these values for each example as well as average values and standard deviations.

\subsection{Results}

\subsubsection{Correctness and Completeness}

After applying our approach to the dataset described above, Tables~\ref{tab:resultsCorrectModels} and~\ref{tab:resultsmutatedmodels} present, for each domain, the number of A, PA, CPA, M, PM and CPM as well as their sum, average and standard deviation at the bottom. They also present precision, recall and F$_1$-measure for each domain, as well as the weighted precision, recall and F$_1$-measure at the bottom of the table. % with respect to the number of A, M or A+M as appropriate at the bottom.
Furthermore, \chg{Table~\ref{tab:resultsCorrectModels}}{Table~\ref{tab:executionTimes}} also shows the execution time it took to run our approach \ins{both broken down by component as well as the total time}.

%%%%%

\begin{table*}[]
\centering
\scriptsize
\caption{Results for Correct Models}
\label{tab:resultsCorrectModels}
%\hskip-0.5cm
\makebox[1 \textwidth][c]{       %centering table
\resizebox{1.3 \textwidth}{!}{   %resize table
\begin{tabular}{c|r|r|r|r|r|r|r|r|r|r|r|r|r|r|r|r}
\multicolumn{1}{l|}{}     & \multicolumn{1}{l|}{A}    & \multicolumn{1}{l|}{PA} & \multicolumn{1}{l|}{CPA} & \multicolumn{1}{l|}{M} & \multicolumn{1}{l|}{PM} & \multicolumn{1}{l|}{CPM} & \multicolumn{1}{l|}{Prec$_a$} & \multicolumn{1}{l|}{Prec$_m$} & \multicolumn{1}{l|}{Prec} & \multicolumn{1}{l|}{Rec$_a$} & \multicolumn{1}{l|}{Rec} & F1$_a$ & F1$_m$ & F1 \\ \hline
R1 & 32 & 24 & 24 & 0 & 0 & 0 & 1.00 & - & 1.00 & 0.75 & 0.75 & 0.86 & - & 0.86 \\ 
R2 & 24 & 23 & 23 & 0 & 0 & 0 & 1.00 & - & 1.00 & 0.96 & 0.96 & 0.98 & - & 0.98 \\ 
R3 & 28 & 21 & 21 & 0 & 0 & 0 & 1.00 & - & 1.00 & 0.75 & 0.75 & 0.86 & - & 0.86 \\ 
R4 & 24 & 14 & 14 & 0 & 2 & 0 & 1.00 & 0.00 & 0.88 & 0.58 & 0.58 & 0.73 & 0.00 & 0.70 \\ 
R5 & 14 & 11 & 11 & 0 & 0 & 0 & 1.00 & - & 1.00 & 0.79 & 0.79 & 0.88 & - & 0.88 \\ 
R6 & 18 & 17 & 17 & 0 & 0 & 0 & 1.00 & - & 1.00 & 0.94 & 0.94 & 0.97 & - & 0.97 \\ 
R7 & 23 & 18 & 18 & 0 & 0 & 0 & 1.00 & - & 1.00 & 0.78 & 0.78 & 0.88 & - & 0.88 \\ 
R8 & 17 & 16 & 16 & 0 & 0 & 0 & 1.00 & - & 1.00 & 0.94 & 0.94 & 0.97 & - & 0.97 \\ 
R9 & 16 & 11 & 11 & 0 & 0 & 0 & 1.00 & - & 1.00 & 0.69 & 0.69 & 0.81 & - & 0.81 \\ 
R10 & 11 & 11 & 11 & 0 & 0 & 0 & 1.00 & - & 1.00 & 1.00 & 1.00 & 1.00 & - & 1.00 \\ 
R11 & 7 & 7 & 7 & 0 & 0 & 0 & 1.00 & - & 1.00 & 1.00 & 1.00 & 1.00 & - & 1.00 \\ 
R12 & 15 & 11 & 11 & 0 & 0 & 0 & 1.00 & - & 1.00 & 0.73 & 0.73 & 0.85 & - & 0.85 \\ 
R13 & 17 & 12 & 12 & 0 & 0 & 0 & 1.00 & - & 1.00 & 0.71 & 0.71 & 0.83 & - & 0.83 \\ 
R14 & 10 & 7 & 7 & 0 & 0 & 0 & 1.00 & - & 1.00 & 0.70 & 0.70 & 0.82 & - & 0.82 \\ 
R15 & 12 & 8 & 8 & 0 & 0 & 0 & 1.00 & - & 1.00 & 0.67 & 0.67 & 0.80 & - & 0.80 \\ 
R16 & 8 & 4 & 4 & 0 & 0 & 0 & 1.00 & - & 1.00 & 0.50 & 0.50 & 0.67 & - & 0.67 \\ 
R17 & 14 & 11 & 11 & 0 & 0 & 0 & 1.00 & - & 1.00 & 0.79 & 0.79 & 0.88 & - & 0.88 \\ 
R18 & 12 & 10 & 10 & 0 & 0 & 0 & 1.00 & - & 1.00 & 0.83 & 0.83 & 0.91 & - & 0.91 \\ 
R19 & 33 & 21 & 21 & 0 & 0 & 0 & 1.00 & - & 1.00 & 0.64 & 0.64 & 0.78 & - & 0.78 \\ 
R20 & 3 & 3 & 3 & 0 & 0 & 0 & 1.00 & - & 1.00 & 1.00 & 1.00 & 1.00 & - & 1.00 \\ 
R21 & 14 & 12 & 12 & 0 & 0 & 0 & 1.00 & - & 1.00 & 0.86 & 0.86 & 0.92 & - & 0.92 \\ 
R22 & 18 & 7 & 7 & 0 & 0 & 0 & 1.00 & - & 1.00 & 0.39 & 0.39 & 0.56 & - & 0.56 \\ 
R23 & 8 & 7 & 7 & 0 & 0 & 0 & 1.00 & - & 1.00 & 0.88 & 0.88 & 0.93 & - & 0.93 \\ 
R24 & 5 & 2 & 2 & 0 & 0 & 0 & 1.00 & - & 1.00 & 0.40 & 0.40 & 0.57 & - & 0.57 \\ 
R25 & 10 & 8 & 8 & 0 & 0 & 0 & 1.00 & - & 1.00 & 0.80 & 0.80 & 0.89 & - & 0.89 \\ 
R26 & 13 & 12 & 12 & 0 & 0 & 0 & 1.00 & - & 1.00 & 0.92 & 0.92 & 0.96 & - & 0.96 \\ 
R27 & 7 & 6 & 6 & 0 & 0 & 0 & 1.00 & - & 1.00 & 0.86 & 0.86 & 0.92 & - & 0.92 \\ 
R28 & 11 & 11 & 11 & 0 & 0 & 0 & 1.00 & - & 1.00 & 1.00 & 1.00 & 1.00 & - & 1.00 \\ 
R29 & 11 & 8 & 8 & 0 & 0 & 0 & 1.00 & - & 1.00 & 0.73 & 0.73 & 0.84 & - & 0.84 \\ 
R30 & 8 & 4 & 4 & 0 & 0 & 0 & 1.00 & - & 1.00 & 0.50 & 0.50 & 0.67 & - & 0.67 \\ \hline
Sum & 443 & 337 & 337 & 0 & 2 & 0 &  &  &  &  &  &  &  & \\
Avg. &  &  &  &  &  &  & 1.00 & 0.00 & \textbf{0.996} & 0.76 & \textbf{0.77} & 0.86 & - & \textbf{0.86} \\
Std. &  &  &  &  &  &  & 0.00 & - & \textbf{0.023} & 0.15 & \textbf{0.17} & 0.12 & - & \textbf{0.12} \\
\end{tabular}
}
}
\end{table*}

As reported in Table~\ref{tab:resultsCorrectModels}, our approach achieves a precision for classifying alignments (Prec$_a$) of 1 for all the domains. This means that when our approach predicts an alignment, it is always correct. The recall (Recall$_a$) is $0.76\pm0.15$, meaning that out of all the alignments, we identify around 76\%. This results in a F$_1$-measure$_a$ of $0.86\pm0.12$.

Regarding the misalignments, the fact that we are running our approach on a set of correct models means that the number of misalignments (M) is 0 (therefore, the Recall$_m$ is undefined). For all the domains, except for R4, the number of predicted misalignments was 0, meaning that our approach did not make any mistake and incorrectly predicted a misalignment that does not exist (i.e., no true negatives). R4 is the only domain model for which our approach incorrectly predicted two misalignments. As a result, Prec$_m$ is undefined for all the domains except R4 where it is 0. Accordingly, the F$_1$-measure$_m$ can only be calculated for R4 and it is 0.

Overall, considering both alignments and misalignments, for the set of correct models, our approach has obtained a precision of $0,996\pm0,023$, a recall of $0.77\pm0.17$ and a F$_1$-measure of $0.86\pm0.12$.

%Let us dig into these two problems were related to the multiplicity of an association end. 
%
%\lola{Not sure whether we need the two following detailed explanations or they should be commented out}
We investigated the two cases in which our algorithm misclassified a correct element as incorrect. Both cases occurred in \textit{R4} and were related to the multiplicities of association ends. We will discuss those two cases in the detailed analysis subsection.

\begin{table*}[]
\centering
\scriptsize
\caption{Results for mutated models (approx. 20\% of model elements are incorrect)}
\label{tab:resultsmutatedmodels}
\makebox[1 \textwidth][c]{       %centering table
\resizebox{1.3 \textwidth}{!}{   %resize table
\begin{tabular}{c|r|r|r|r|r|r|r|r|r|r|r|r|r|r|r}
\multicolumn{1}{l|}{}     & \multicolumn{1}{l|}{A}    & \multicolumn{1}{l|}{PA} & \multicolumn{1}{l|}{CPA} & \multicolumn{1}{l|}{M} & \multicolumn{1}{l|}{PM} & \multicolumn{1}{l|}{CPM} & \multicolumn{1}{l|}{Prec$_a$} & \multicolumn{1}{l|}{Prec$_m$} & \multicolumn{1}{l|}{Prec} & \multicolumn{1}{l|}{Rec$_a$} & \multicolumn{1}{l|}{Rec$_m$} & \multicolumn{1}{l|}{Rec} & F1$_a$ & F1$_m$ & F1 \\ \hline
R1 & 31 & 25 & 25 & 5 & 4 & 4 & 1.00 & 1.00 & 1.00 & 0.81 & 0.80 & 0.81 & 0.89 & 0.89 & 0.89 \\
R2 & 20 & 20 & 20 & 4 & 4 & 4 & 1.00 & 1.00 & 1.00 & 1.00 & 1.00 & 1.00 & 1.00 & 1.00 & 1.00 \\
R3 & 24 & 17 & 17 & 3 & 2 & 2 & 1.00 & 1.00 & 1.00 & 0.71 & 0.67 & 0.70 & 0.83 & 0.80 & 0.83 \\
R4 & 15 & 10 & 10 & 6 & 4 & 4 & 1.00 & 1.00 & 1.00 & 0.67 & 0.67 & 0.67 & 0.80 & 0.80 & 0.80 \\
R6 & 16 & 14 & 14 & 2 & 2 & 2 & 1.00 & 1.00 & 1.00 & 0.88 & 1.00 & 0.89 & 0.93 & 1.00 & 0.94 \\
R7 & 15 & 11 & 11 & 5 & 0 & 0 & 1.00 & - & 1.00 & 0.73 & 0.00 & 0.55 & 0.85 & - & 0.71 \\
R8 & 11 & 10 & 10 & 5 & 2 & 2 & 1.00 & 1.00 & 1.00 & 0.91 & 0.40 & 0.75 & 0.95 & 0.57 & 0.86 \\
R9 & 7 & 6 & 6 & 5 & 3 & 3 & 1.00 & 1.00 & 1.00 & 0.86 & 0.60 & 0.75 & 0.92 & 0.75 & 0.86 \\
R10 & 8 & 8 & 8 & 2 & 2 & 2 & 1.00 & 1.00 & 1.00 & 1.00 & 1.00 & 1.00 & 1.00 & 1.00 & 1.00 \\
R11 & 7 & 7 & 7 & 1 & 1 & 1 & 1.00 & 1.00 & 1.00 & 1.00 & 1.00 & 1.00 & 1.00 & 1.00 & 1.00 \\
R12 & 13 & 9 & 9 & 1 & 1 & 1 & 1.00 & 1.00 & 1.00 & 0.69 & 1.00 & 0.71 & 0.82 & 1.00 & 0.83 \\
R13 & 10 & 8 & 8 & 4 & 2 & 2 & 1.00 & 1.00 & 1.00 & 0.80 & 0.50 & 0.71 & 0.89 & 0.67 & 0.83 \\
R14 & 4 & 3 & 3 & 4 & 3 & 3 & 1.00 & 1.00 & 1.00 & 0.75 & 0.75 & 0.75 & 0.86 & 0.86 & 0.86 \\
R15 & 10 & 7 & 7 & 2 & 1 & 1 & 1.00 & 1.00 & 1.00 & 0.70 & 0.50 & 0.67 & 0.82 & 0.67 & 0.80 \\
R16 & 6 & 3 & 3 & 2 & 1 & 1 & 1.00 & 1.00 & 1.00 & 0.50 & 0.50 & 0.50 & 0.67 & 0.67 & 0.67 \\
R17 & 9 & 8 & 8 & 4 & 4 & 4 & 1.00 & 1.00 & 1.00 & 0.89 & 1.00 & 0.92 & 0.94 & 1.00 & 0.96 \\
R18 & 10 & 8 & 8 & 2 & 1 & 1 & 1.00 & 1.00 & 1.00 & 0.80 & 0.50 & 0.75 & 0.89 & 0.67 & 0.86 \\
R19 & 27 & 16 & 16 & 6 & 5 & 5 & 1.00 & 1.00 & 1.00 & 0.59 & 0.83 & 0.64 & 0.74 & 0.91 & 0.78 \\
R20 & 2 & 2 & 2 & 1 & 1 & 1 & 1.00 & 1.00 & 1.00 & 1.00 & 1.00 & 1.00 & 1.00 & 1.00 & 1.00 \\
R21 & 13 & 12 & 12 & 1 & 0 & 0 & 1.00 & - & 1.00 & 0.92 & 0.00 & 0.86 & 0.96 & - & 0.92 \\
R22 & 16 & 6 & 6 & 2 & 1 & 1 & 1.00 & 1.00 & 1.00 & 0.38 & 0.50 & 0.39 & 0.55 & 0.67 & 0.56 \\
R23 & 8 & 7 & 7 & 1 & 0 & 0 & 1.00 & - & 1.00 & 0.88 & 0.00 & 0.78 & 0.93 & - & 0.88 \\
R25 & 8 & 7 & 7 & 2 & 2 & 2 & 1.00 & 1.00 & 1.00 & 0.88 & 1.00 & 0.90 & 0.93 & 1.00 & 0.95 \\
R26 & 11 & 10 & 10 & 2 & 2 & 2 & 1.00 & 1.00 & 1.00 & 0.91 & 1.00 & 0.92 & 0.95 & 1.00 & 0.96 \\
R27 & 7 & 6 & 6 & 2 & 2 & 2 & 1.00 & 1.00 & 1.00 & 0.86 & 1.00 & 0.89 & 0.92 & 1.00 & 0.94 \\
R28 & 8 & 8 & 8 & 3 & 2 & 2 & 1.00 & 1.00 & 1.00 & 1.00 & 0.67 & 0.91 & 1.00 & 0.80 & 0.95 \\
R29 & 10 & 8 & 8 & 1 & 1 & 1 & 1.00 & 1.00 & 1.00 & 0.80 & 1.00 & 0.82 & 0.89 & 1.00 & 0.90 \\
R30 & 6 & 3 & 3 & 1 & 1 & 1 & 1.00 & 1.00 & 1.00 & 0.50 & 1.00 & 0.57 & 0.67 & 1.00 & 0.73 \\ \hline
Sum & 301 & 234 & 234 & 74 & 50 & 50 &  &  &  &  &  &  &  &  & \\
Avg. &  &  &  &  &  &  & 1.00 & 1.00 & \textbf{1.00} & 0.78 & 0.68 & \textbf{0.77} & 0.88 & 0.87 & \textbf{0.87}\\
Std. &  &  &  &  &  &  & 0.00 & 0.00 & \textbf{0.00} & 0.16 & 0.29 & \textbf{0.16} & 0.11 & 0.15 & \textbf{0.11}\\
\end{tabular}
} % end resize
} % end centering
\end{table*}

Table~\ref{tab:resultsmutatedmodels} presents the results for the dataset with mutated domain models, containing approximately 20\% of wrong model elements\footnote{Note that we could not introduce mutations in the models of R5 and R24 because they would not produce incorrect models w.r.t. the specification.}.  

For all the models the Precision$_a$ is 1, which means that all the predicted alignments are correct. The Recall$_a$ is $0.78\pm0.16$, which means that our approach made correct predictions for around 78\% of the alignments. The F$_1$-measure$_a$ is $0.88\pm0.11$. It is worth noting that the difficulties with R4 in the first experiment that only contained correct model elements disappeared. This is due to the fact that, coincidentally, the two associations causing the algorithm to make an incorrect prediction were mutated.

Precision$_m$ is also 1 for all the cases of mutated domain models, except for R7, R21 and R23 in which it is undefined because our approach did not predict any misalignment correctly (the denominator is 0). Recall$_m$ is $0.68\pm0.29$, which means that our approach detected correctly around 68\% of the misalignments and the F$_1$-measure$_m$ is $0.87\pm0.15$.

Finally, overall, the Precision is always 1, the Recall is $0.77\pm0.16$ and the F$_1$-measure is $0.87\pm0.11$.

\subsubsection{Detailed Analysis for each Model Element Type}
\label{sec:AnalysisPerModelElementType}

In this subsection, we study whether there is a significant difference in performance of our algorithm for different model element types. We do so by revisiting the results for each category of model element. For space reasons we only present the summary of the analysis, but the detailed tables listing our results per model element category can be found in our repository~\cite{repo}.

\paragraph{Attributes}

Our algorithm achieves a precision of 1 for attributes. The recall is 1.0 for 15 out of 30 models, the average recall is 0.83, and the lowest recall, 0.33, appears in two models: \me{Cinema} and \me{Communication Pref}. The algorithm displays several consistent patterns of failing to verify the alignment of attributes.
In some cases, the attribute is not explicitly mentioned in the textual description, or the noun in the textual description differs from the attribute name in the class diagram. For instance, in the \me{R21-Savings Account domain} example, the attribute \me{balance} is not explicitly stated, but it is implicitly understood that a \me{savings account} class must have a \me{balance} attribute.
In seven cases, the problem was that the LLM applied temporal reasoning. \ins{However, time is typically ignored in domain models, especially when specifying attributes.} For example, if the \me{User} class contains an \me{email} attribute \ins{of type \me{String}}, the sentence generated by the \emph{Rule-based Model Sentence Generator} (component D)  is ``A user has an email.''. In R21, the textual description states: ``If the user has not any e-mail address registered with the banking system, the e-mail option should display as ‘disabled,’ and by default, the SMS option is selected.'' Here, the user may not possess an email\ins{, or might have initially registered with an email and later removed it}. In such scenarios, the LLM \del{used}indicated that more context is necessary to draw a definite conclusion and \ins{our algorithm} outputs \textit{Not Sure} after the equivalence, contradiction and inclusion checks.

\paragraph{Association Relationships}

For identifying alignments and misalignments of association relationships and multiplicities, the algorithm achieved an average precision of 0.995, an average recall of 0.76, and an average F1-score of 0.865.

As reported already, in the \me{R4-Galaxy Sleuth Game} domain, two \emph{aligned} associations were wrongly predicted as \emph{misaligned} by the algorithm. We thoroughly investigated these two cases. Both cases were related to the multiplicities of an association end. In the first case, the textual requirements talked about \textit{two associations between the same two classes}. When classifying one of the associations, our approach confused it with the other one: the LLM detected an incorrect multiplicity because it compared the generated sentence with a sentence from the original text that was describing \ins{the} other association.
In the second case, the model contained an association between the class \me{Player} and the class \me{Hypothesis} with multiplicity \texttt{*}. The requirements stated that ``If an announced hypothesis is incorrect, the player [...] cannot pose hypotheses any longer [...]''. The LLM applied \insTwo{again the} \textit{temporal reasoning} \insTwo{pattern} and concluded that, while the multiplicity between \me{Player} and \me{Hypothesis} should be \texttt{*} \textit{before} the player makes an incorrect hypothesis, \textit{after} making a wrong hypothesis the multiplicity should be 0. Based on this reasoning the LLM concluded that the multiplicity in the model was incorrect.

It is also worth mentioning that a closer inspection of the textual requirements of R4 -- a description of a board game -- revealed that the text is very vague and ambiguous. For example, the requirements refer to the same concepts with different names interchangeably (e.g., \textit{hypothesis} and \textit{question}), which not only affects the matching done by the rule-based semantic matcher component, but could also affect the understanding of the LLM. Of course, real-world requirements sometimes do contain synonyms, despite the fact that most rigorous requirements elicitation techniques try to eliminate them.

Since the average recall is 0.78, we also looked into the cases where the LLM was not able to classify the model element to identify patterns. In 25 cases, the generated sentences were ambiguous because \textit{the modeler did not provide role names for the association}. When a role name is missing, we use \me{has} as a verb to form the sentence. For example, for the association from the \me{Section} to the \me{Restaurant} class with multiplicity \me{1}, the generated sentence is ``A section has a restaurant.'' This is ambiguous because a \me{restaurant} has \me{sections}, not the other way around.

In six cases, the \textit{Semantic Matcher} component matched the wrong sentence from the textual description, which led the LLM to return an ``inconclusive'' response. In two cases, the \textit{spaCy }library classified role names as nouns instead of verbs, producing grammatically incorrect sentences. For example, for the association from the \me{Rental} class to the \me{Bike} class with the role name \me{rents} and multiplicity \me{1}, the generated sentence was ``A Rental has rents which is a bike''. This sentence is incorrect. If \textit{spaCy} had correctly identified ``rents'' as a verb, the sentence would have been ``A rental rents a bike'', which is a correct sentence and would have been classified as aligned by the LLM.

In two other cases, the generated sentences were judged as not precise enough by the LLM when the association had a multiplicity \me{0..*} because we used the words ``can have'' to build the sentence. For example, for the association from the \me{User} class to the \me{Reservation} class with multiplicity \me{0..*}, the generated sentence was ``A user can have reservations''. Here, multiplicity \me{0..*} means a user may have no reservations, but the LLM was not sure whether ``can have'' also includes ``may have no''.

\paragraph{Composition Relationships}

For composition relationships, the algorithm achieved an average precision of 1, an average recall of 0.58, and an average F1-score of 0.49. Since the recall when identifying alignments in compositions is low, we analyzed the LLM responses to understand the reasons. In seven cases, the LLM interpreted that the composed class must be a physical part of the composite class, and if this was not true, it could not reach a clear conclusion. For example, in the \me{ATM} example (R14), there is a composition from the \me{Bank} class to the \me{Account} class. The relevant description states, ``Each bank provides its computer to maintain its accounts and process transactions against them.'' The generated sentence is ``Each bank is made up of accounts'' but since a \me{Bank} is not physically made up of \me{accounts}, the LLM could not identify this as an alignment.
In two other cases, the composition was implicit and spread across multiple sentences in the textual description, making it difficult to determine the relationship from a single sentence.
%In the remaining two cases, the semantic matcher did not match the relevant sentence from the description, which led to incorrect results. 

\paragraph{Generalization/Specialization Relationships}

For inheritance relationships, the algorithm achieved an average precision of 1 and an average recall of 0.85. The algorithm did not achieve a recall of 1.0 for three models: \me{R2 Employee Management}, \me{R5 Spy-Robot Game}, and \me{R29 Prepaid Card}. In the \me{Employee Management} domain, it missed the alignment between \me{Off-Shift Worker} and \me{Worker} classes. This occurred because the textual description used the term ``non-shift worker'' instead of ``off-shift worker'', making it difficult for the algorithm to confirm the alignment. In the \me{Spy-Robot Game} domain, the algorithm did not detect the alignment between the \me{Enemy} and \me{Spaceship} classes since the relationship was stated indirectly in the sentence that describes how to play the game: `` `Jump' actions are used to dodge the aliens on the ground and the empty spaces made by the bombs dropped by the spaceships''. The LLM did, however, detect the generalization/specialization relationship between \me{Enemy} and \me{Alien}, which is more directly expressed in the sentence. Finally, for the \me{Prepaid Card} domain, the \textit{Semantic Matcher} component did not select the correct sentence from the description.

\paragraph{Enumerations}

For enumerations, the algorithm achieves a precision of 1, \del{and} an average recall of 0.72 and an average F1-score of 0.81. The recall is 1.0 for 6 out of 16 models, while the lowest recall of 0.33 appears for the \me{Library} example. We identified several recurrent problems in failing to verify alignment for enumerations.

In four cases, the enumeration literal is not present in the textual description. For example, in the \me{Library} domain, the \me{Level} enumeration has three enumeration literals: \me{Beginner, Intermediate} and \me{Advanced}. In the textual description, it is mentioned that ``A language tape has a title language (e.g., Spanish) and level (e.g., beginner)''. In the text, only \me{Beginner} is mentioned as a level. In two of the cases, the enumeration is not informative. For example, in the \me{Library} domain, the \me{Status} enum has three enumeration literals: \me{Borrowed, Reserved and Renewed}. The sentence in the textual description says ``An item can be borrowed, reserved, or renewed to extend a current loan''. Here, the LLM is not sure whether the \me{Status} attribute actually refers to the loan status of the item or not. We experimented with renaming the enumeration type to \me{Loan Status}, and then the LLM detected an alignment.  Also, in two of the cases, the enumeration literals were not explicitly mentioned, but were mentioned as part of some other action. For example, in the \me{Supermarket} domain, the \me{Order Status} enumeration has two enumeration literals: \me{Started} and \me{Completed}. The sentence in the description says ``The delivery process begins when the customer first interacts with the service organization and ends when the delivery of the desired service is completed and the customer exits the process''. Here, ``started'' is not mentioned at all, and ``completed'' is mentioned implicitly. % Also in two of the cases, \me{Other} is used as an enumeration literal, and in the description it is not mentioned anywhere.

\paragraph{Summary}

\ins{The detailed analysis for each model element type revealed interesting facts about our algorithm. First, the only mistake our algorithm made was when classifying the multiplicities of association ends. For all other model elements, the precision is 100\%. This suggests that our algorithm can even be used in a completely automated setting for model elements that are not association ends.}

\ins{Second, about half of the cases where the algorithm was unable to classify the model element fall into some patterns (e.g., the application of temporal reasoning as explained above). Hence there is an opportunity for further research to determine whether it is possible to teach the LLM how to overcome these recurring classification difficulties, e.g., by using few-shot prompting or fine tuning.}

%%%%%%%%

\subsection{Scalability}
\label{sec:ResultsScalability}
To study the scalability of our approach, we first discuss the complexity of the classification algorithm illustrated with pseudocode in Algorithm~\ref{alg:algoritm}. The only operation that has a significant performance cost is the querying of the LLM, which happens at least 3 and at most 19 times \textit{per model element} (outer \texttt{for loop} lines~\ref{line:alg1_begin_for}-\ref{line:alg1_end_for} and \textit{per matched sentence} (inner \texttt{for loop} lines \ref{line:alg1_begin_inner_for}-\ref{line:alg1_end_inner_for})\footnote{5 times for equivalence checks, 9 times for contradiction, and 7 times for inclusion}. If we denote the number of model elements with $m$ and the number of sentences in the textual specification with $s$, then the worst-case complexity of our algorithm is O($m s$). This is, however, a very pessimistic worst case, as it assumes that the sentence matcher component matches a model element with \textit{every} sentence. In our general experience, which was also confirmed in our experiment, this only happens with small textual requirements. As the textual requirements grow, different paragraphs and sections talk about different concepts, which allows the sentence matcher to avoid matching model elements with unrelated sentences.

\ins{Fortunately, the queries associated with each model element are mutually independent. Even for a given model element, each question is independent of the other ones. In fact, new sessions with the LLM are started for each query on purpose so that so that the LLM answers each questions based on its inherent knowledge of the domain alone, without being influenced by previous questions. Hence it is actually possible to run \textit{all LLM questions in parallel}, which in theory reduces the cost of querying the LLM to a constant: the maximum time it takes \textit{one} LLM query to complete. Whether in practice the queries actually are evaluated by the LLM in parallel or not depends on the LLM and the user conditions. In our experiments with GPT-4o, we initially noticed some throttling happening at the LLM, because of the token and query limits of our OpenAI account. After using the account for a while, though, we got upgraded to a higher usage tier. At the time of running our experiment, our account had reached usage tier 4, which accepts 10k requests per minute and 2M tokens per minute.}

\begin{table}[]
\centering
\scriptsize
\caption{\ins{Component Execution Times (in seconds)}}
\label{tab:executionTimes}
\begin{tabular}{l|rr|rr|rr|rr|r|r|r}
    & \multicolumn{2}{c|}{\textbf{A}}                        & \multicolumn{2}{c|}{\textbf{C}}                      & \multicolumn{2}{c|}{\textbf{D}}                      & \multicolumn{2}{c|}{\textbf{E}}                      & \multicolumn{1}{c|}{\textbf{Total}} & \multicolumn{1}{c|}{\textbf{min}} & \multicolumn{1}{c}{\textbf{max}} \\
    & \multicolumn{1}{c}{Total} & \multicolumn{1}{c|}{/word} & \multicolumn{1}{c}{Total} & \multicolumn{1}{c|}{/me} & \multicolumn{1}{c}{Total} & \multicolumn{1}{c|}{/me} & \multicolumn{1}{c}{Total} & \multicolumn{1}{c|}{/me} & \multicolumn{1}{c|}{}               & \multicolumn{1}{c|}{\textbf{/me}} & \multicolumn{1}{c}{\textbf{/me}} \\ \hline
R1  & 115.89                    & 0.45                       & 135.74                    & 4.242                    & 6.93                      & \textbf{0.22}            & 35.93                     & 1.12                     & 04:54                               & 0:16                              & 0:40                             \\
R2  & 141.65                    & 0.62                       & 9.06                      & 0.377                    & 9.25                      & 0.39                     & 11.79                     & \textbf{0.49}            & 02:52                               & 0:08                              & 0:12                             \\
R3  & 117.61                    & 0.56                       & 13.24                     & 0.473                    & 9.33                      & 0.33                     & \textbf{103.38}           & 3.69                     & 04:04                               & 0:14                              & \textbf{1:43}                    \\
R4  & \textbf{657.65}           & \textbf{0.93}              & 58.12                     & 2.422                    & 6.92                      & 0.29                     & 52.99                     & 2.21                     & \textbf{12:56}                      & \textbf{0:27}                     & 0:55                             \\
R5  & 286.97                    & 0.70                       & 7.80                      & 0.557                    & 9.35                      & 0.67                     & 25.02                     & 1.79                     & 05:29                               & 0:15                              & 0:26                             \\
R6  & 65.38                     & 0.51                       & 7.92                      & 0.440                    & 4.40                      & 0.24                     & 12.96                     & 0.72                     & 01:31                               & 0:07                              & 0:13                             \\
R7  & 335.06                    & 0.62                       & 47.70                     & 2.074                    & \textbf{11.06}            & 0.48                     & 42.59                     & 1.85                     & 07:16                               & 0:20                              & 0:44                             \\
R8  & 158.03                    & 0.51                       & 0.23                      & 0.013                    & 7.90                      & 0.46                     & 36.13                     & 2.13                     & 03:22                               & 0:19                              & 0:36                             \\
R9  & 370.80                    & 0.86                       & 0.21                      & 0.013                    & 7.74                      & 0.48                     & 30.78                     & 1.92                     & 06:50                               & 0:16                              & 0:30                             \\
R10 & \textbf{41.71}            & 0.50                       & 0.12                      & 0.011                    & 7.15                      & 0.65                     & 9.86                      & 0.90                     & \textbf{00:59}                      & \textbf{0:05}                     & 0:10                             \\
R11 & 48.51                     & 0.36                       & 0.05                      & 0.007                    & 5.00                      & 0.71                     & 9.26                      & 1.32                     & 01:03                               & \textbf{0:05}                     & 0:10                             \\
R12 & 95.71                     & 0.80                       & 19.63                     & 1.308                    & 4.86                      & 0.32                     & 16.65                     & 1.11                     & 02:17                               & 0:08                              & 0:18                             \\
R13 & 167.70                    & 0.72                       & 11.54                     & 0.679                    & 7.11                      & 0.42                     & 29.68                     & 1.75                     & 03:36                               & 0:15                              & 0:30                             \\
R14 & 118.15                    & 0.72                       & 14.36                     & 1.436                    & 5.85                      & 0.59                     & 10.29                     & 1.03                     & 02:29                               & 0:08                              & 0:12                             \\
R15 & 115.75                    & 0.52                       & 3.96                      & 0.330                    & 4.33                      & 0.36                     & 20.26                     & 1.69                     & 02:24                               & 0:10                              & 0:20                             \\
R16 & 107.51                    & 0.63                       & 11.36                     & 1.420                    & 3.92                      & 0.49                     & 12.97                     & 1.62                     & 02:16                               & 0:09                              & 0:14                             \\
R17 & 47.80                     & 0.50                       & 0.13                      & 0.010                    & 4.13                      & 0.30                     & 14.57                     & 1.04                     & 01:07                               & 0:07                              & 0:14                             \\
R18 & 121.10                    & 0.82                       & 0.07                      & 0.006                    & 4.32                      & 0.36                     & 25.91                     & 2.16                     & 02:31                               & 0:12                              & 0:26                             \\
R19 & 208.20                    & 0.69                       & \textbf{203.34}           & \textbf{6.162}           & 7.85                      & 0.24                     & 63.66                     & 1.93                     & 08:03                               & 0:16                              & 1:09                             \\
R20 & 59.60                     & 0.65                       & 0.64                      & 0.212                    & \textbf{3.34}             & \textbf{1.11}            & \textbf{7.54}             & 2.51                     & 01:11                               & 0:06                              & \textbf{0:08}                    \\
R21 & 111.19                    & 0.29                       & 0.08                      & 0.006                    & 6.60                      & 0.47                     & 14.74                     & 1.05                     & 02:13                               & 0:09                              & 0:15                             \\
R22 & 147.58                    & 0.36                       & 4.05                      & 0.225                    & 5.83                      & 0.32                     & 36.68                     & 2.04                     & 03:14                               & 0:18                              & 0:37                             \\
R23 & 145.30                    & 0.34                       & 0.05                      & 0.006                    & 5.12                      & 0.64                     & 11.90                     & 1.49                     & 02:42                               & 0:08                              & 0:12                             \\
R24 & 103.18                    & 0.33                       & \textbf{0.03}             & 0.005                    & 4.03                      & 0.81                     & 8.43                      & 1.69                     & 01:56                               & 0:07                              & 0:19                             \\
R25 & 159.95                    & 0.53                       & 1.50                      & 0.150                    & 6.05                      & 0.61                     & 14.70                     & 1.47                     & 03:02                               & 0:07                              & 0:15                             \\
R26 & 81.86                     & \textbf{0.28}              & 0.07                      & 0.006                    & 7.47                      & 0.57                     & 13.38                     & 1.03                     & 01:43                               & 0:08                              & 0:13                             \\
R27 & 179.43                    & 0.36                       & 0.13                      & 0.019                    & 4.45                      & 0.64                     & 29.20                     & \textbf{4.17}            & 03:33                               & 0:16                              & 0:29                             \\
R28 & 128.90                    & 0.36                       & 0.14                      & 0.013                    & 4.08                      & 0.37                     & 30.82                     & 2.80                     & 02:44                               & 0:16                              & 0:30                             \\
R29 & 138.50                    & 0.41                       & 0.06                      & 0.006                    & 5.66                      & 0.51                     & 13.24                     & 1.20                     & 02:37                               & 0:08                              & 0:13                             \\
R30 & 77.08                     & 0.31                       & \textbf{0.03}             & \textbf{0.003}           & 4.53                      & 0.57                     & 10.44                     & 1.30                     & 01:32                               & 0:06                              & 0:10                            
\end{tabular}
\end{table}

\ins{
Table~\ref{tab:executionTimes} shows the execution times for each component when running the experiment with the correct models\footnote{The running times for the models containing errors are identical, because the number and kind of model elements are the same, and because of the parallelization the inclusion questions are also asked even if the answers of the LLM in the end are not being used for elements that are classified as misaligned.}. The most time is spent in component A, the \textit{NLP Specification Preprocessor}, which took 42s for the smallest model (R10 with 83 words), and more than 11 minutes for the largest model (R4 with 710 words). Per word, the fastest pre-processing took place in R26 (0.28s per word), and the longest in R4 (0.93s per word).}

\ins{The processing time of component B, the \textit{Model Slider}, is negligible, as for the biggest model it took in total less than 0.6ms. Hence these times are omitted from Table~\ref{tab:executionTimes}. The performance of component C, the \textit{Semantic Matcher}, varies significantly, depending on whether matches are found immediately or whether more elaborate comparison is needed. For several models it computed the matching sentences quasi instantly (0.003s per model element for R30), whereas it took 6.162s per model element for R19. In general, though, the matching is fast, with a median of 0.180s per model element.
Component D, the \textit{Sentence Generator}, took minimally 0.22s per model element in R1, and maximally 1.11s in R19. Finally, component E, the \textit{Alignment Detector}, thanks to the parallel execution of the prompts, the times range from 7.54s for the fastest model (R20) to 1 minute and 43s for the longest one (R3). Per model element, the fastest processing was R2 with 0.49s and the slowest R28 with 4.17s.
%However, for component E, these averages are only theoretical, since the processing runs in parallel.
}

\ins{In the end, the fastest model to be processed \textit{in its entirety} took 59 seconds. That model, R10, contains 11 model elements and its description has 83 words. The longest model to process was R4 with 12 minutes and 56 seconds. R4 has 24 model elements and the description is comprised of 710 words. These processing times include the preprocessing step.}

\ins{In case our approach is used to process \textit{one model element}, it makes sense to look at the measured times by 1) excluding the preprocessing time of component A, as it needs to be run one time only, 2) summing up the average time it takes components B, C and D to process a model element, and 3) using the minimum or maximum processing times that it takes component E to completely process the prompts of one model element. These times are reported in the last two columns in Table~\ref{tab:executionTimes}. We see that the fastest processing times of 5s per model element were recorded for models R10 and R11, whereas the slowest processing time of 1min 43s was recorded during the processing of R3.}

\del{However, even if the complexity is not exponential, the actual running time is considerable. We measured the time it took to run the experiments for correct models (see last column in Table~\ref{tab:resultsCorrectModels})}\footnote{\del{The running time for models with errors would be faster, since the algorithm does not check for inclusion if the contradiction check is positive}}. \del{The fastest model was analyzed in a little more than 5 minutes, whereas the biggest model took approximately. The fastest classification time for a single model element was 18 seconds, and the longest took 1 minute and 3 seconds.}

%%%%%%%%

\subsection{Answers to the Research Questions}
\hfill

\textbf{Answer to RQ1.} To answer RQ1 about correctness, we inspect the overall measured precision, shown on the last row in bold in Table~\ref{tab:resultsCorrectModels} and Table~\ref{tab:resultsmutatedmodels}. On correct models (i.e., models without misalignments), our approach made predictions on 339 model elements, and only misclassified 2 as misaligned (precision of 0.996). On more realistic models, in our case models containing a minimum of 20\% of model elements with mistakes, \textit{all our predictions} were correct (precision of 1.0). In other words, if our approach makes a prediction, the prediction is almost certainly correct.

\textbf{Answer to RQ2.} To answer RQ2 about completeness, we turn to the measured recall. As one can see from the last row in Table~\ref{tab:resultsCorrectModels} and Table~\ref{tab:resultsmutatedmodels}, the average recall is 0.77 and 0.78, which means that on average we were able to make correct predictions for slightly more than 3 out of 4 model elements.

\textbf{Answer to RQ3.} To answer \ins{R3} about scalability, we have discussed the complexity of our approach and have reported on the execution times. In the best-case scenario the complexity of our approach is linear and in the worst-case scenario it is quadratic\del{(i.e., it is not exponential), the real-world execution time remains significant}. In our experiments, checking a single model element takes between \chg{18}{5} seconds to 1 minute \ins{43s}, whereas applying the approach to an entire model takes \chg{5 minutes}{59s} for the smallest model, while the largest model required \chg{2 hours}{12m 56s}.

\subsection{Threats to Validity}

\subsubsection{Conclusion Validity}
%Conclusion validity affects the ability to draw correct conclusion about the relations between the treatment and the outcome, i.e., how reasonable the conclusion is. Examples that influence this threat to validity include the choice of sample size, and the measurement of the experiment.

The fact that LLMs are by nature non-deterministic~\cite{song2024goodbadgreedyevaluation} is another threat to conclusion validity. To mitigate this problem, %for most model elements,
we prompt the LLM and ask about the same pair of model element and matching sentence several times with different questions.

\subsubsection{Construct Validity}
%Construct validity refers to the extent to which the experiment setting reflects the theory, i.e., whether the research/tests are well-constructed using established standards and methods. For example, whether the type of samples are representative of the population or not; or whether the number of classes taken reflects common experience.
\del{The correctness and completeness of our approach have been studied using a third-party benchmark dataset. Although the size of the requirements and models in the dataset are relatively small, they cover different and diverse domains as well as a large part of the concepts that can be found in domain models. Because of the heterogeneity of the domain models in the dataset, we had to ask a colleague to either create some domain models in the form of class diagrams or to convert the structured text representing a domain model into a class diagram. This person might have had a biased view of the domains.}
%Threats to construct validity relate to how alignment quality is operationalized and measured.

\ins{In this study, alignment quality is quantified using precision, recall, and F1, which may not fully capture the relevance or practical usefulness of alignments. For instance, all alignments are treated as equally important, despite the fact that they might have different relevance and impact. To shed more light on the practical usefulness of our approach, we plan to carry out an empirical study with human participants in the future.}

%Even if our evaluation has been done with a dataset that does not contain large requirements, dataset designed for a university course, we have selected advanced specifications and models on which the modeling capabilities of students were evaluated. We believe that these exercises are representative, in terms of complexity, of any domain model that a professional modeler could face.

\subsubsection{Internal Validity}
% Internal validity checks whether the test or instrument measures what it is supposed to. This threat can affect the independent variable with respect to causality. That is, the results may indicate a causal relationship, although there is none.
\del{There are several aspects that would affect the effectiveness of our approach considerably, such as a high number of alignments and especially misalignments that are not predicted, or the incorrect prediction of (mis)alignments.}

\del{To study these aspects, in our experiments, we have used precision and recall to identify these issues. We report on its results in detail.
We are also reporting the execution times of our experiments. However, for this aspect, a more thorough performance study (e.g., repeated executions, etc.) is required to be able to draw more meaningful conclusions.}

\ins{
%Threats to internal validity concern factors that may influence the observed results independent of the proposed approach.
There are several aspects that could affect the results presented in this paper.
The correctness and completeness of our approach have been studied using a third-party dataset~\cite{chaudron2024}. % consisting of pairs of textual descriptions and corresponding domain models, from where the ground-truth alignments were taken. 
Although the size of the requirements and models in the dataset are relatively small, they cover different and diverse domains as well as a large part of the concepts that can be found in domain models. 
While the use of a third-party dataset reduces subjectivity and bias, we assumed the models derived from the descriptions to be correct. Any errors, inaccuracies or missing elements in the models may have affected our evaluation. Furthermore, because of the heterogeneity of the domain models in the dataset, we required the help of a human to create some models and transform structured text into class diagrams. The alignments derived from these artifacts may reflect subjective judgments.
%This is construct validity---> Finally, the inherent non-determinism of LLM outputs and researcher choices regarding model and parameter selection may introduce variability in the results. To mitigate this we used several prompts for each test (namely, equivalence, contradiction and inclusion) and computed the majority vote.
}

\subsubsection{External Validity}
% This kind of threat limits the ability to generalize the results beyond the experiment context.
The use of a third-party dataset covering diverse domains and the fact that it includes positive and negative examples (i.e., correct models and models with errors) and models with different sizes gives us some confidence that our results can be generalized to other cases. However, we cannot generalize our results for large specifications and domain models as we know there is some performance penalty when the size of these grows and further research in needed in this line. Finally, we only used one LLM hence we cannot generalize our results to other LLMs.
\vspace{-1mm}
\section{Discussion on Applicability}
\label{sec:discussion}

The evaluation results for our approach are very promising, but since the precision is not 100\% 
%not all kind of errors can be detected, 
and the running time is non-negligible, thought must be given on how to best integrate the approach into model-driven development. This section revisits the results, puts them in perspective and, if applicable, provides some ideas on how the algorithm could be improved.

\subsection{Correctness}

The correctness of the algorithm is excellent, with a precision of 1 for all models and all model elements except for R4, where two multiplicies of associations were incorrectly classified as misaligned. As a result, our approach could be used effectively within a modeling tool \textit{as a modeling assistant}: model elements classified as aligned by our approach would be highlighted as ``verified correct'', whereas model elements classified as misaligned could be highlighted as ``suspected incorrect''. In a follow-up step we would run more experiments to confirm that the classification of our algorithm is 100\% correct for the following model element categories: attributes, inheritance relationships and enumerations. If this is confirmed, then the algorithm could even be used for their automated validation.

However, because our misalignment precision is not 100\%, we imagine that our approach would need to allow the modeler to consult the sentence in the textual specification that contradicts what was modeled, as well as read the reasoning provided by the LLM for flagging the model element as misaligned. The modeler can then decide on their own whether indeed the model element is misaligned, or whether to discard the warning of the assistant.

\subsection{Completeness}

The recall for all model element categories combined is currently above 0.75, which means that the modeler would see confirmation indications for more than 75\% of their model elements if our approach is used as part of an assistant. We believe that it might be possible to improve this result further by specifically targeting the situations that we identified in our experiment in which our approach was not able to provide a classification.

For associations, the most common reason for not being able to provide a classification is that the modeler did not provide role names. Forcing the modeler to put role names on association ends would increase recall considerably.

In many other cases, the LLM was not able to provide a clear answer because domain models usually do not model time constraints, whereas sentences in problem descriptions often describe situations that evolve over time. For example, a garage might provide services during weekdays, but not on weekends. In these situations, the LLM would apply temporal reasoning and doubt whether the sentence in the problem description ``A garage provides services to customers during weekdays, e.g., changing the oil or the tires.'' aligns with the sentence ``A garage can provide services'' (which implies ``a garage can \textit{always} provide services'') generated from the domain model. It would be interesting to explore whether the prompts could be modified (or augmented with few shot prompting) to instruct the LLM to not apply temporal reasoning.

\subsection{Scalability}
\label{sec:ApplicabilityScalability}

In practice, domain modeling is a relatively slow and iterative activity~\cite{Ali25}. We believe that the delay introduced by our approach should not affect the experience too negatively when used as part of a modeling assistant that runs in the background. \ins{In such a scenario, the preprocessing component A would run only once on the problem description. Components B to E would then produce misalignment annotations as they become available, always progressively calculated on single model elements as the modeler adds them to the model. With this low volume of calls to the LLM, we would be expecting processing times close to the fastest measured times, i.e., below 10s per model element.} \del{, producing alignment or misalignment annotations as they become available, always progressively calculated on single model elements and on not an entire model.} \ins{This would be similar to background compilation or test execution in an IDE, which also produces warnings or test results incrementally as they become available.}

%As already mentioned in section~\ref{sec:motivation},
We could also imagine our approach being used in an offline mode, for example, by running overnight to establish traceability links between textual requirements and domain models, or tag the model with a quality estimate that can be used to decide on whether human inspection is required or not. \ins{In this case, where speed is not essential, one could even imagine running component E in sequential mode. This would save LLM cost (i.e., energy and money), because as soon as a model element can be classified because the relative majority for a prompt has been reached, the remaining queries for that model element would not have to be sent to the LLM. These savings can be considerable. For example, running the error-free R17 model in parallel mode resulted in costs of US \$5.82, whereas running it in sequential mode only cost US \$1.79. However, while the parallel execution time for R17 was 1m 7s, the sequential execution time was 24m 44s.}

Finally, it remains to be seen whether our approach is effective on textual requirements and models that are orders of magnitude bigger. Since the cost of querying the LLM is significant, the performance of the algorithm could be improved significantly by using local LLMs as they become more available \ins{or smaller LLMs (i.e., LLMs with less parameters such as GPT-4o mini)}. \ins{Even if local or smaller LLMs have limited reasoning capabilities compared to their larger counterparts, ongoing research and advanced fine-tuning methods are significantly enhancing their performance on complex tasks~\cite{tian2024,bucher2024}.} \del{On the other hand, while so far the NLP processing of the textual specification and the sentence matching time was \chg{negligible}{much lower}, we would have to investigate whether this is still the case when the textual specifications get significantly larger.}
\ins{We would also have to investigate the performance cost of the NLP preprocessing step and sentence matcher component as the textual specifications get significantly larger. Since the precision of component E is very high, we could explore replacing those components by a much simpler matcher.}

\subsection{\ins{Modeler Experience}}
\label{sub:ModelerExperience}

\ins{Since our approach can be integrated in an assistant, the requirements and expectations from novice modelers and expert modelers might differ. For instance, empirical evidence~\cite{BenChaabenBDS25} shows that less experienced modelers often welcome stronger and more explicit guidance, whereas experts typically favor subtle, on-demand support. The integration of the approach into an assistant should take these aspects into account and offer different models of assistance.}

\subsection{Limitations}

Our approach is currently not able to detect certain errors, e.g., using the wrong type when defining attributes. Missing or unnecessary model elements can also not be detected. Similarly, misplaced attributes can also not be detected, as they are treated as additional ones. Therefore, in order to be able to detect the full spectrum of mistakes, a modeling assistant would have to combine our approach with some of the existing techniques mentioned in the related work. 

%\lola{We might want to say here that we our contribution here is the approach, not how the approach will be used. For instance, it could be used in runtime (i.e., when the model is being built), or on the final model (i.e., when the modeler thinks the model is finished).}
\section{Related Work}
\label{sec:rw}

\chg{We situate our work with respect to the literature on (semi-)automated domain modeling and the evaluation of domain models, as both deal with the identification of links between domain models and textual specifications.} {In this paper, we use NLP and LLMs to verify the alignment between domain model elements and textual specifications. While no prior work directly addresses this specific verification task, several research areas provide relevant context: (semi-)automated domain modeling approaches, LLM applications in domain modeling, schema and ontology matching techniques, and domain model evaluation methods. However, these areas differ from our approach in their objectives, methods, or scope, as we detail below.}

\subsection{(Semi-)Automated Domain modeling}
\label{sec:semi_automated}
\del{Manual construction of domain models from textual specifications is a time-consuming, labor-intensive, and error-prone task. Existing research addresses these challenges by employing rules, machine learning, or deep learning to assist modelers in completing partial domain models through recommendations}~\cite{weyssow2022recommending,di2021gnn,burgueno2021nlp, di2023memorec}\del{, or by extracting domain models from textual specifications}~\cite{yang2022towards,saini2022machine,robeer2016automated, francuu2011automated}\del{. For instance, Weyssow et al.}~\cite{weyssow2022recommending}\del{ fine-tuned a pre-trained RoBERTa}~\cite{liu2019robertarobustlyoptimizedbert}\del{ language model to recommend meaningful domain concepts relevant to the given modeling context. Burgueño et al.}~\cite{burgueno2021nlp}\del{ segmented the partial domain model into attributes and relationships, and used a pretrained language model to recommend domain concepts for model completion. Rocco et al.}~\cite{di2023memorec}\del{ provided recommendations by considering the existing portion of the metamodel as active context, using a graph representation to encode relationships among metamodel artifacts and generating recommendations with a context-aware collaborative filtering technique}~\cite{schafer2007collaborative}\del{. In the area of extracting domain models from textual descriptions, our previous approach}~\cite{saini2022machine}\del{ introduced a hybrid method that combines rule-based and machine learning methods to extract domain models from given specifications. This method uses the Spacy library along with smaller machine learning models trained for specific tasks. Similarly, Yang et al.}~\cite{yang2022towards}\del{ utilized NLP rules and task-specific machine learning models. Both approaches evaluate the domain model extracted by the machine learning model against a predefined `golden truth' model. However, unlike other methods, our earlier work}~\cite{saini2022machine}\del{, recognizes that a textual description can be modeled in multiple valid ways. To address this, the approach uses an interactive mode to engage with the user and decide how a particular user requirement should be modeled, such as choosing between using the Player-role pattern, using inheritance or using enumerations. This enables modeling the user requirements in multiple ways.}

\ins{Manual construction of domain models from textual specifications is time-consuming and error-prone. Existing research addresses these challenges through recommendations or extraction approaches.} 

\ins{Recommendation-based approaches assist modelers complete partial domain models. For instance, Weyssow et al.~\cite{weyssow2022recommending} fine-tuned a pre-trained RoBERTa~\cite{liu2019robertarobustlyoptimizedbert} language model to recommend meaningful domain concepts relevant to the given modeling context. Burgueño et al.~\cite{burgueno2021nlp} segmented the partial domain model into attributes and relationships, and used a pretrained language model to recommend domain concepts for model completion. Rocco et al.~\cite{di2023memorec} provided recommendations by considering the existing portion of the metamodel as active context, using a graph representation to encode relationships among metamodel artifacts and generating recommendations with a context-aware collaborative filtering technique~\cite{schafer2007collaborative}.}

\ins{Extraction-based approaches generate domain models from textual descriptions. Yang et al.~\cite{yang2022towards} and Saini et al.~\cite{saini2022machine} use a hybrid approach by combining NLP rules with task-specific machine learning models to extract model elements, evaluating outputs against reference solutions. Both approaches evaluate the domain model extracted by the machine learning model against a predefined `golden truth' model. Unlike other methods, Saini et al. acknowledged multiple valid modeling solutions for the same textual description and incorporated interactive user feedback for modeling decisions}

\ins{In contrast to these approaches, our proposed method in this paper does not aim to automatically extract domain models from textual specifications or to recommend new model elements to the modeler. Instead, we evaluate the alignment of the model elements that the modeler uses in their partial domain model. We classify these elements as correct or incorrect using the textual domain specification as the `golden truth'.}

\subsection{LLMs for Domain modeling}
As LLMs continue to advance, their benefits are becoming more apparent in various aspects of the development processes. Consequently, model-based engineering (MDE) is increasingly integrating diverse LLMs for a range of modeling tasks, including automated domain modeling~\cite{chen2023}. Several studies have incorporated LLMs to extract domain model elements from given domain specifications, without any human interaction or traditional supervised training on a specific domain or task~\cite{10.1145/3652620.3687807, prokop2024enhancing, chaaben2023towards}.

% As large language models (LLMs) continue to advance, their applications for providing automation and/or assistance during software development are becoming increasingly effective. Not surprisingly, the model-driven engineering (MDE) community is also integrating LLMs in various different ways to perform a range of modeling tasks, including automated domain modeling~\cite{chen2023}. Recent studies employ LLMs to extract domain model elements from domain specifications without human involvement or traditional supervised training for specific domains or tasks ~\cite{10.1145/3652620.3687807, prokop2024enhancing, chaaben2023towards}. They are discussed in the following paragraphs.

For instance, Chen et al. ~\cite{chen2023} conduct a comparative study on the use of LLMs for fully automated domain modeling. They assess GPT-3.5 and GPT-4, employing various prompt engineering techniques such as zero-shot, N-shot, and Chain-of-Thought (CoT) prompting on a dataset containing ten diverse domain modeling examples with reference solutions created by modeling experts. They experiment with five cases: one zero-shot prompt, two 1-shot prompts, one 2-shot prompt, and one prompt with CoT reasoning. Their findings show that, while LLMs demonstrate strong domain understanding capabilities, they remain impractical for full automation. The top-performing GPT-4 achieves F1 scores of 0.76 for class generation, 0.61 for attribute generation, and 0.34 for relationship generation. Moreover, the F1-score exhibits higher precision and lower recall, so LLM-retrieved domain elements are often reliable, but many elements are missing. Additionally they observe that generated domain models rarely follow modeling best practices. Furthermore, adding examples to the prompt improves LLM performance in retrieving classes and relations, but including reasoning steps can also decrease performance.

The same authors extend their research in~\cite{10.1145/3652620.3687807} by proposing a multi-step automated domain modeling approach that extracts model elements from problem descriptions using GPT-4. Their approach includes step-by-step instructions and follows an iterative process. They provide a task description to outline the overall domain model generation task, a natural language modeling problem description to specify the problem domain, and an example store with few-shot examples that define model elements and patterns for the multi-step iterative generator (MIG).

Compared to the average results of their previous single-step approach, the multi-step method improves the F1-score for class identification by 22.71\% (from 0.6280 to 0.7706) and for relationship identification by 75.18\% (from 0.1781 to 0.3120). The F1-score for attributes remains similar, while the F1-score for identifying the Player-Role pattern improves by 10.39\%.

Prokop et al.~\cite{prokop2024enhancing} explore the automation of domain modeling using pre-trained large language models (LLMs) by presenting an experimental LLM-based conceptual modeling assistant that collaborates with a human expert. The assistant offers modeling suggestions based on a given textual description of the domain. They use zero-shot and N-shot prompting techniques. Given a textual description and a partial model M, they first query the LLM to generate all classes. For each class, they query the LLM for attributes, and in the next step, they query for associations. A human expert evaluates each suggestion provided by the LLM before integrating it into the domain model. They experiment with the Mistral-7B and Llama-3 8B models. Their experiments show that Llama generally performs better than Mistral, but Mistral with Retrieval-Augmented Generation (RAG) outperforms Llama without RAG.

Chaaben et al.~\cite{chaaben2023towards} address the completion of partial models using a few-shot prompting technique with the GPT-3 model. Their approach supports both static and dynamic diagram completion. They begin by semantically mapping the domain concepts from the partial model into a structured text representation. In the next step, they query the LLM using this structured representation along with a set of few-shot examples to generate relevant domain concepts.

The recommendation process proceeds in multiple steps. First, they ask the LLM to suggest class names. Next, for each class, they prompt the LLM to suggest attributes, providing only the attributes from the partial model in the input prompt. Finally, they query for associations between each pair of classes, again using only the relevant information from the partial model in the prompt for that specific association. Their results show a precision of 0.56 and a recall of 0.45 for class name generation, a recall of 0.7 for attribute suggestions, and an accuracy of 0.64 for identifying associations.

\chg{Unlike these approaches, which use LLMs to process entire textual requirements, our method partitions the model into small model slices. These approaches use LLMs either to generate class diagrams from textual requirements or to recommend relevant model elements, whereas our approach employs LLMs to compare pairs of sentences, asking for simple yes/no answers. Our method reduces the complexity faced by the LLM, as it only needs to semantically compare two sentences.}{Unlike these approaches that process entire textual requirements and use LLMs to generate or recommend relevant model elements, our method partitions the model into small model slices. In addition, our method uses LLMs to compare pairs of sentences, asking for simple yes/no answers. Our method reduces the complexity faced by the LLM, as it only needs to semantically compare two sentences.} Our process is fully automated and does not require any human interaction. Additionally, we use zero-shot prompting, eliminating the need to design domain-agnostic sample examples for few-shot prompt settings.

Unlike these approaches that process entire requirements to generate or recommend elements, our method partitions models into small slices and uses LLMs to perform semantic comparisons between specification sentences and model element descriptions. This reduces LLM complexity, requires no human interaction, and uses zero-shot prompting without domain-specific examples.

% Added whole section
\subsection{Schema and Ontology Matching}
\label{sec:RW_schemaAndOntologyMatching}
\ins{
Schema and ontology matching-based methods address the challenge of identifying semantic correspondences between heterogeneous data structures. While these methods share conceptual similarities with our work in terms of semantic alignment, the problem formulation, artifacts involved, and verification objectives differ substantially.}

\ins{The ontology matching literature distinguishes between two primary methodological approaches. Schema-based matching techniques rely on structural and terminological features of the ontology, such as class labels, property names, and hierarchical relationships between concepts~\cite{sharma2022lsmatch, faria2022matcha}. In contrast, instance-based matching methods use entity-level information, including the specific classes and properties instantiated for individual entities, to infer semantic correspondences~\cite{ayala2021towards, belhadi2020data, fallatah2022impact}. Both approaches aim to establish mappings between elements of two distinct ontological structures.}

\ins{Recent work has explored the integration of large language models and pre-trained embeddings to enhance semantic matching capabilities. For instance, Hao et al. propose a system that introduces heterogeneous graph layers to incorporate both the local structure and the global context into concept embeddings using SBERT embedding method~\cite{hao2021medto}. Wang proposes a knowledge graph embedding technique for filtering the candidates generated by a language model. Other approaches use pre-trained SBERT model for encoding labels~\cite{kossack2022tom, peng2023ontology}. Ahmed et al. propose prompting LLMs with fragments of the source and target metamodels, identifying correspondences through an iterative process. The fragments to be provided in the prompt are identified based on an initial mapping derived from their elements’ definitions~\cite{11025595}.}

\ins{The above approaches differ from our approach in several fundamental aspects. First, ontology matching addresses a correspondence identification problem between two independently constructed and equally valid schemas or metamodels, i.e., a symmetric model-to-model alignment task. In contrast, our work formulates an asymmetric verification problem where domain model elements are evaluated against textual specifications that serve as the ground truth reference. Second, while ontology matching seeks to discover and establish mappings between corresponding elements across two complete structural representations, our approach performs correctness assessment of individual model elements relative to textual specification fragments, yielding binary classification outcomes rather than correspondence sets. These distinctions highlight that while both areas involve semantic alignment, they address fundamentally different problems with distinct inputs, objectives, and evaluation criteria.}

\subsection{Evaluation of Domain Models}
Evaluating domain models for their coherence and conciseness is a time consuming task. Several approaches propose methods to evaluate domain models in comparison to a `golden truth' domain model that represents a reference solution. Bian et al.~\cite{8904595} propose a grading algorithm to match student solutions with the template solution using syntactic, semantic, and structural matching strategies. Boubekeur et al.~\cite{10.1145/3417990.3418741} propose an approach that uses simple heuristics and machine learning techniques to evaluate and categorize student submissions of domain models based on their quality.

As already mention in subsection~\ref{sec:semi_automated}, Yang and Sahraoui~\cite{yang2022towards} propose an approach to generate domain models automatically using natural language patterns and machine learning. The authors then evaluate these generated models with reference solutions using exact
matching, relaxed matching, and general matching over classes
and relationships.

Singh et al.~\cite{10.1145/3550356.3561583} develop a mistake detection system (MDS) to identify different types of mistake, such as missing elements, extra elements, incorrect elements, and incorrect application of modeling patterns. The MDS compares a
student solution with a correct solution provided by the instructor by iterating over all the elements in both solutions and performing various checks. Kasaei et al.~\cite{10892739} propose an approach for equivalence checking to detect semantically similar concepts within domain models. Their approach uses the combination of pre-trained model and dictionary to identify equivalent elements.

In contrast, our approach does not require a reference \ins{model} solution, because it considers the given textual specification as the `golden truth' for comparison and evaluation. However, our approach can currently only detect errors with respect to wrong model elements. Missing or unnecessary model elements are currently not detected.

\vspace{-1mm}
\section{Conclusions}
\label{sec:conclusions}
\vspace{-1mm}

In this paper, we presented an approach to detect alignments and misalignments between a textual specification written in natural language and an existing, potentially partial domain model expressed as a UML class diagram. Our approach first uses NLP techniques to preprocess the textual specification. Next, we split the domain model into minimal model slices focusing on one particular model element. Finally, we generate a sentence in natural language for each of the model elements in focus, and query an LLM to discover alignments or misalignments between this sentence and selected sentences from the textual specification that talk about that model element. Ultimately, our approach classifies every model element as \textit{aligned} with the textual specification, \textit{misaligned} with the textual specification, or leaves it \textit{unclassified}. 

We evaluated our approach on a publicly available set of 30 requirements specifications from a broad range of domains. Our results show nearly perfect precision for detecting alignments and misalignments. Moreover, we achieve an average recall of 77\% for alignments and 78\% for misalignments.

We identified several ways in which our approach could be extended. First, we could investigate whether it is possible to refine the prompts for dealing with multiplicities to avoid misclassifications when there are multiple associations between the same two classes, as well as to avoid that the LLM interprets temporal constraints expressed in the text as contradictions. Additionally, \ins{since our detailed analysis of the LLM responses detected several patterns that caused the LLM to not be able to classify model elements with enough certainty, we are planning to explore few-shot prompting and/or fine-tuning techniques to train the LLM to recognize these patterns and avoid them. Finally, }we want to \chg{integrate}{extend our approach to be able to handle} additional model elements.
%investigate whether our approach can also be adapted to additional model elements.

Regarding scalability, we are planning to investigate ways of speeding up the execution, e.g., using a locally hosted and fine-tuned LLM, as well as experimenting with significantly larger requirements and models. Finally, we would like to use our approach to implement an assistant for a modeling tool in order to be able to perform a user study.

%Future work:\\
%- Include more model elements\\
%- Validate with further examples\\
%- Apply to different types of models\\
%- Build an assistant and perform an empirical study to study the usability\\

\section*{Acknowledgement}
This project has been funded by the Spanish Ministry of Science, Innovation, and Universities under contract PID2021-125527NB-I00 and Universidad de Málaga under project JA.B1-17 PPRO-B1-2023-037.

%\appendix
%\section{Example Appendix Section}
%\label{app1}


%%       \cite{<label>} ==> [1]

%%




\bibliographystyle{elsarticle-num} 
\bibliography{literature}

\end{document}